\documentclass[12pt]{article}
\usepackage{amssymb,amsmath,epsfig}
\allowdisplaybreaks
\begin{document}
\title{\bf Models of Collapsing and Expanding Cylindrical Source in $f(R,T)$ Theory}
\author{M. Sharif \thanks{msharif.math@pu.edu.pk} and Aisha Siddiqa
\thanks{aisha.siddiqa17@yahoo.com}\\
Department of Mathematics, University of the Punjab,\\
Quaid-e-Azam Campus, Lahore-54590, Pakistan.}

\date{}

\maketitle
\begin{abstract}
We discuss the collapsing and expanding solutions of anisotropic
charged cylinder in the context of $f(R,T)$ theory ($R$ represents
the Ricci scalar and $T$ denotes the trace of energy-momentum
tensor). For this purpose, we take an auxiliary solution of
Einstein-Maxwell field equations and evaluate expansion scalar whose
negative values lead to collapse and positive values give expansion.
For both cases, the behavior of density, pressure, anisotropic
parameter as well as mass is explored and the effects of charge as
well as model parameter on these quantities are examined. The energy
conditions are found to be satisfied for both solutions.
\end{abstract}

\section{Introduction}

Gravitational force is amenable for governing many astrophysical
phenomena like formation of stars, keeping stars together in
galaxies, gravitational collapse and restricting the heavenly bodies
in their respective orbits. A star is in equilibrium state under the
balance of pressure (directed outward) and gravity (directed
inward). It undergoes collapse if gravity exceeds pressure and
experiences expansion when pressure overcomes gravity. During the
life, a star experiences both these phenomena. Oppenheimer and
Snyder \cite{1a} are pioneers to study the gravitational collapse
for dust matter. Misner and Sharp worked on the collapse of a star
considering isotropic \cite{1b} as well as anisotropic fluid
\cite{1c}.

After that many researchers studied the process of collapse for
different configurations. Stark and Piran \cite{1d3} examined the
gravitational waves emitted by the gravitational collapse of
rotating relativistic polytropes. Herrera \textit{et al}. \cite{1d1}
examined the dynamical instability of spherical symmetric collapsing
fluid suffering heat dissipation and showed that dissipation
increases the instability. Harada \cite{1d2} investigated the final
outcome of gravitational collapse of a sphere with perfect fluid
distribution and discussed the limits when the singularity is naked
or not. Joshi and Dwivedi \cite{1d4} explored the final outcome of
spherical symmetric dust collapse. Depending upon the initial
pressure and density distribution, they discussed different new
black hole solutions.

The $f(R)$ gravity is a direct generalization of general relativity
(GR) obtained by replacing $R$ with $f(R)$ in the Einstein-Hilbert
action. Many astrophysical and cosmological phenomena have been
explored within the scenario of $f(R)$ theory. The contributions of
$f(R)$ terms can lead different consequences on various phenomena.
Sharif and Kausar \cite{2a} discussed the perfect fluid collapse in
this theory and solved the equations assuming constant Ricci scalar.
They showed that $f(R)$ terms play the role of anti gravitational
force. Cembranos {\it et al.} \cite{2b} worked on spherical dust
collapse and showed that the contribution of $f(R)$ terms slows down
the collapsing process. Hence in general, it can be deduced that the
inclusion of higher order curvature terms reduce the collapse rate.
Also in GR, a gravitational wave has two polarization modes while in
$f(R)$ theory it is shown that gravitational wave has two extra
modes than GR \cite{6c}.

Harko \textit{et al}. \cite{4a} proposed a more generalized
gravitational theory known as $f(R,T)$ gravity. The curvature-matter
coupling produces a deviation from geodesic motion which may yield
interesting results and help to explore dark side of the universe
\cite{4b}. Shabani and Farhoudi \cite{4c} investigated cosmological
viability of some $f(R,T)$ gravity models using solar system
constraints. Moraes \textit{et al.} \cite{4c1} studied the
equilibrium configurations of neutron and quark stars in this theory
concluding that mass can cross observational limits. Noureen and
Zubair \cite{5e} investigated the stability of anisotropic spherical
star in the framework of $f(R,T)$ yielding some constraints on
physical quantities. Carvalho \textit{et al.} \cite{4c2} analyzed
white dwarfs using an Equation of state describing ionized atoms
embedded in a relativistic Fermi gas of electrons in
curvature-matter coupling scenario. They observed that white dwarfs
have larger radius and mass in $f(R,T)$ gravity than those observed
in GR and $f(R)$ theory.

Recent detection of gravitational waves has brought motivation to
study the collapse phenomenon with the existence of gravitational
waves in the exterior. It is well-known by Bhirkoff's theorem that a
spherical symmetric vacuum spacetime cannot have gravitational
radiation. In this context, the next assumption is cylindrical
system, because Einstein and Rosen found exact solution of the field
equations which models the propagation of cylindrical gravitational
waves. Sharif and Bhatti \cite{5a} discussed charged expansion-free
cylindrical system and found that stability is controlled by charge,
density and pressures. Yousaf \textit{et al.} \cite{5h} discussed
the stability of cylindrical stellar system through perturbation
technique and found its dependence on the stiffness parameter,
matter variables as well as $f(R,T)$ dark source terms. Sharif and
Farooq investigated the dynamics of charged cylindrical collapse in
$f(R)$ gravity with perfect \cite{5b} as well as bulk viscous
dissipative fluid \cite{5c} and concluded that collapse rate slows
down due to dark source terms.

Rosseland and Eddington \cite{3a} were the first who figured out the
possibility that stars can have electric charge. After that the
presence of electromagnetic field in self-gravitating systems is
explored by many researchers. The effect of charge on spherical
collapse \cite{3c1} and on the stability of compact objects
\cite{3d} have been investigated with the conclusion that charge
halts the collapse as well as increases the stability regions.
Bhatti and Yousaf \cite{7d} explored the effects of electromagnetic
field on plane symmetric anisotropic dissipative fluid configuration
in Palatini $f(R)$ gravity. They concluded that matter inhomogeneity
is enhanced with charge while it is decreased due to modified
gravity terms. Mansour \textit{et al.} \cite{7f} analyzed the
features of compact stars in the presence of weak electromagnetic
field in $f(R)$ gravity.

Glass \cite{8a} studied the collapsing and expanding solutions for a
non-static anisotropic spherical source within the scenario of GR.
Abbas extended this work for plane symmetric configuration
\cite{8b}, charged spherical source \cite{8c} and for charged
cylindrical geometry \cite{8d} in GR as well as for sphere in
$f(R,T)$ gravity \cite{8e}. We discussed these solutions for charged
spherical configuration in $f(R,T)$ theory \cite{8f}. In this paper,
we investigate the effects of charge on the evolution of a
non-static cylindrical source in $f(R,T)$ gravity. The paper is
planned as follows. In the coming section, we discuss the outline of
work done then in next one we formulate the Einstein-Maxwell
equations for $f(R,T)=R+2\lambda T$ gravity model (where $\lambda$
is coupling constant also called model parameter) and discuss the
cases of collapse and expansion. In the last section, we summarize
our results.

\section{Physical Goals}

In this section, we first discuss physical goals of the research
work presented in this paper and then elaborate the technique to
achieve these objectives. Here, we would like to explore physical
characteristics of a cylindrical star during the phases of collapse
and expansion in the dark energy dominated era. When a star starts
to loose the hydrostatic equilibrium, firstly, its outer layers
expand and it becomes a red-giant. However, after some time, the
star suffers a supernova explosion and experiences a collapse. In
order to discuss the whole scenario in an expanding universe, we
consider $f(R,T)$ theory of gravity as discussed in the introduction
as an alternative to GR. Also, to extend our discussion, we observe
the effects of electromagnetic field and consider a charged star. We
aim to discuss the behavior of density, pressures, pressure
anisotropy and mass of the star as well as check the energy
conditions for the obtained solutions. We also investigate the
effects of curvature-matter coupling and presence of charge on the
collapse and expansion phases of the star's life.

For this purpose, we generate collapsing and expanding solutions for
our cylindrically symmetric model. We then analyze the physical
parameters graphically such that the density and mass remain
positive. Also, the obtained values of density and pressures must
satisfy the energy conditions for the viability of the solution
otherwise there is a possibility of existence of exotic fluid that
is an unrealistic situation. The value of curvature-matter coupling
constant is taken such that our model $f(R,T)=R+2\lambda T$
satisfies the viability conditions
\begin{equation}\nonumber
f_{R}>0,\quad 1+\frac{f_{T}}{8\pi}>0 \quad \text{and}\quad f_{RR}>0,
\end{equation}
which yield the constraint $\lambda>-4\pi$ for the considered model.
In the graphical analysis, free parameters appearing in the solution
are fixed such that our solution is physically acceptable, i.e.,
mass as well as density are positive and energy conditions are
satisfied. In order to examine the effects of electromagnetic field
and curvature-matter coupling constant, we vary the values of the
total charge and $\lambda$ in the plots and check the corresponding
increase and decrease in the respective quantity.

\section{Einstein-Maxwell Field Equations}

The $f(R,T)$ gravity action with the contribution of electromagnetic
field is defined as
\begin{equation}\label{1}
S=\int d^{4}x\sqrt{-g}\left[\frac{1}{16\pi}f(R,
T)+\mathcal{L}_{m}+\mathcal{L}_{e}\right].
\end{equation}
The electromagnetic Lagrangian density $\mathcal{L}_{e}$ has the
form $\mathcal{L}_{e}=mF_{\mu\nu}F^{\mu\nu},~m$ is an arbitrary
constant, $F_{\mu\nu}=\phi_{\nu,\mu}-\phi_{\mu,\nu}$ represents the
electromagnetic field tensor and $\phi_{\mu}$ represents the four
potential. The field equations for the above action are
\begin{equation}\label{2}
f_{R}R_{\mu\nu}-\frac{1}{2}g_{\mu\nu}f+(g_{\mu\nu}\Box-
\nabla_{\mu}\nabla_{\nu})f_{R}=8\pi T_{\mu\nu}
-f_{T}(T_{\mu\nu}+\Theta_{\mu\nu})+8\pi E_{\mu\nu},
\end{equation}
where $f_{R}$ and $f_{T}$ denote the derivatives of $f(R,T)$ with
respect to $R$ and $T$, respectively. The expression for
$\Theta_{\mu\nu}$ is given by
\begin{equation}\label{3}
\Theta_{\mu\nu}=g^{\gamma\alpha}\frac{\delta
T_{\gamma\alpha}}{\delta g^{\mu\nu}}, \quad
T_{\mu\nu}=g_{\mu\nu}\mathcal{L}_{m}- \frac{\partial
\mathcal{L}_{m}}{\partial g^{\mu\nu}},
\end{equation}
and the electromagnetic energy-momentum tensor $E_{\mu\nu}$ is
defined by
\begin{equation}\label{3a}
E_{\mu\nu}=\frac{1}{4\pi}\left(F_{\mu}^{~\alpha}F_{\nu\alpha}-
\frac{1}{4}F^{\alpha\beta}F_{\alpha\beta}g_{\mu\nu}\right).
\end{equation}
The non-static cylindrically symmetric spacetime is taken as
\begin{equation}\label{6}
ds^{2}=-A^{2}(t,r)dt^{2}+B^{2}(t,r)dr^{2}+C^{2}(t,r)d\theta^{2}+dz^{2}.
\end{equation}
The energy-momentum tensor for anisotropic fluid is given by
\begin{equation}\label{7}
T_{\mu\nu}=(\rho+p_{r})V_{\mu}V_{\nu}+p_{r}g_{\mu\nu}-(p_{r}-p_{z})S_{\mu}S_{\nu}
-(p_{r}-p_{\theta})K_{\mu}K_{\nu},
\end{equation}
where $V_{\mu}$ denotes the four velocity, $S_{\mu}$, $K_{\mu}$ are
unit four-vectors, $\rho$ stands for density, $p_{r}$, $p_{\theta}$,
$p_{z}$ are the pressures in $r$, $\theta$ and $z$ directions,
respectively. The four-vectors $V_{\mu}$, $S_{\mu}$ and $K_{\mu}$
have the expressions
\begin{equation}\label{7a}
V^{\mu}=(A^{-1},0,0,0),\quad K^{\mu}=(0,0,C^{-1},0),\quad
S^{\mu}=(0,0,0,1),
\end{equation}
which satisfy the following relations
\begin{equation}\nonumber
V^{\mu}V_{\nu}=-1,~ K^{\mu}K_{\nu}=S^{\mu}S_{\nu}=1,~
S^{\mu}K_{\nu}=V^{\mu}K_{\nu}=V^{\mu}S_{\nu}=0.
\end{equation}

The Maxwell equations are given by
\begin{equation}\label{3aa}
F^{\mu\nu}_{~~;\nu}=4\pi j^{\mu},
\end{equation}
where $j^{\mu}$ represents the four current. In comoving frame, the
four potential and four current are defined as
\begin{equation}\label{3a3}
\phi_{\mu}=(\phi,0,0,0),\quad j_{\mu}=(\xi,0,0,0),
\end{equation}
$\phi$, $\xi$ (both are functions of $t$ and $r$) represent electric
scalar potential and charge density, respectively. The Maxwell
equations for the metric (\ref{6}) yield
\begin{eqnarray}\label{3b}
\phi^{''}+\phi^{'}\left(\frac{C'}{C}-\frac{A'}{A}-\frac{B'}{B}\right)&=&4\pi
\xi AB^{2},\\\label{3b2}
\dot{\phi}^{'}+\phi^{'}\left(\frac{\dot{C}}{C}-\frac{\dot{A}}{A}-\frac{\dot{B}}{B}\right)&=&0.
\end{eqnarray}
Integration of Eq.(\ref{3b}) gives
\begin{equation}\label{3b1}
\phi^{'}=\frac{AB}{C}q,\quad q=4\pi\int_{0}^{r}\xi BCdr,
\end{equation}
where $q$ is the total charge of the cylinder. We take
$f(R,T)=R+2\lambda T$ proposed by Harko \textit{et al.} \cite{4a} to
explore the effects of curvature-matter coupling on collapsing and
expanding solutions. This model has frequently been used in
literature \cite{r9} which yields a power-law type scale factor and
is able to discuss accelerated expansion of the universe. It
corresponds to $\Lambda$CDM model with trace dependent cosmological
constant or $\Lambda(T)$ gravity discussed by Poplawski \cite{r10}.
The above mentioned expression of $f(R,T)$ and
$\mathcal{L}_{m}=-\rho$ simplify the field equations as
\begin{equation}\label{4a}
G_{\mu\nu}=(8\pi+2\lambda) T_{\mu\nu}+2\lambda \rho
g_{\mu\nu}+\lambda Tg_{\mu\nu}+8\pi E_{\mu\nu},
\end{equation}
which produces the following set of equations
\begin{eqnarray}\label{8}
\frac{1}{B^{2}}\left[\frac{B'C'}{BC}-\frac{C''}{C}\right]+
\frac{1}{A^{2}}\frac{\dot{B}\dot{C}}{BC}+\frac{A^{2}q^{2}}{C^{2}}=
8\pi\rho-\lambda(-\rho+p_{r}+p_{\theta}+p_{z}),
\\\label{9}
\frac{\dot{C}'}{C}-\frac{\dot{C}}{C}\frac{A'}{A}-\frac{\dot{B}}{B}\frac{C'}{C}=0,\\\label{10}
\frac{1}{A^{2}}\left[\frac{\dot{A}C'}{AC}-\frac{\ddot{C}}{C}\right]+
\frac{1}{B^{2}}\frac{A'C'}{AC}+\frac{B^{2}q^{2}}{C^{2}}= 8\pi
p_{r}+\lambda(\rho+3p_{r}+p_{\theta}+p_{z}),\\\nonumber
\frac{1}{A^{2}}\left[\frac{\dot{A}\dot{B}}{AB}-\frac{\ddot{B}}{B}\right]+
\frac{1}{C^{2}}\left(\frac{\dot{C}}{B}\right)^{2}\left[\frac{A''}{A}-\frac{A'B'}{AB}\right]-
\frac{q^{2}}{C^{2}}=8\pi
p_{\theta}\\\label{11}+\lambda(\rho+p_{r}+3p_{\theta}+p_{z}),
\\\nonumber
\frac{1}{A^{2}}\left[-\frac{\ddot{B}}{B}-\frac{\ddot{C}}{C}-
\frac{\dot{B}\dot{C}}{BC}+\frac{\dot{A}}{A}\left(\frac{\dot{B}}{B}+\frac{\dot{C}}{C}\right)\right]
+\frac{1}{B^{2}}\left[\frac{A''}{A}+\frac{C''}{C}-\frac{A'B'}{AB}\right.\\\label{12}\left.-
\frac{C'}{C}\left(\frac{B'}{B}+\frac{A'}{A}\right)\right]-\frac{q^{2}}{C^{2}}=8\pi
p_{z}+\lambda(\rho+p_{r}+p_{\theta}+3p_{z}),
\end{eqnarray}
where dot and prime represent differentiation with respect to $t$
and $r$, respectively.

A simultaneous solution of the field equations give the following
explicit expressions of density and pressure components
\begin{eqnarray}\nonumber
\rho&=&\frac{1}{8(8\pi^{2}+6\pi\lambda+\lambda^{2})A^{3}B^{3}C^{2}}
\left[q^{2}(8\pi+5\lambda)A^{5}B^{3}-2AB^{2}C(-2(2\pi\right.\\\nonumber
&+&\left.\lambda)\dot{B}\dot{C}
+\lambda(C\ddot{B}+B\ddot{C}))+\lambda
B^{2}C\dot{A}(2C\dot{B}+B(\dot{C}+2C'))-\lambda
A^{2}\right.\\\nonumber
&\times&\left.(C^{2}+\dot{C}^{2})(A'B'-BA'')+A^{3}(-2q^{2}\lambda
B^{3}+q^{2}\lambda
B^{5}+4(2\pi+\lambda)\right.\\\label{13}&\times&\left.CB'C'-4(2\pi+\lambda)BCC'')\right],\\\nonumber
p_{r}&=&\frac{1}{8(2\pi+\lambda)(4\pi+\lambda)A^{3}B^{3}C^{2}}\left[-q^{2}\lambda
A^{5}B^{3}+q^{2}A^{3}B^{3}(2\lambda+(8\pi\right.\\\nonumber
&+&\left.3\lambda)B^{2})-2AB^{2}C(-\lambda
C\ddot{B}+(4\pi+\lambda)B\ddot{C})+B^{2}C\dot{A}(-2\lambda
C\dot{B}\right.\\\nonumber
&+&\left.B(-\lambda\dot{C}+2(8\pi+3\lambda)C'))+A^{2}(4(2\pi+\lambda)BCA'C'+\lambda
(C^{2}\right.\\\label{14}&+&\left.\dot{C}^{2})(A'B'-BA''))\right],\\\nonumber
p_{\theta}&=&\frac{1}{8(2\pi+\lambda)(4\pi+\lambda)A^{3}B^{3}q^{2}}\left[-q^{2}\lambda
A^{5}B^{3}-q^{2}A^{3}B^{3}(8\pi+2\lambda\right.\\\nonumber
&+&\left.\lambda B^{2})-2AB^{2}C((4\pi+\lambda)C\ddot{B}-\lambda
B\ddot{C})+B^{2}C\dot{A}(2(4\pi+\lambda)C\dot{B}\right.\\\label{15}&-&\left.\lambda
B(\dot{C}+2C'))+A^{2}(\lambda
C^{2}-(8\pi+3\lambda)\dot{C}^{2})(A'B'-BA'')\right],\\\nonumber
p_{z}&=&\frac{-1}{8(2\pi+\lambda)(4\pi+\lambda)A^{3}B^{3}C^{2}}\left[q^{2}\lambda
A^{5}B^{3}+2AB^{2}C(2(2\pi+\lambda)\dot{B}\dot{C}\right.\\\nonumber
&+& \left.
(4\pi+\lambda)(C\ddot{B}+B\ddot{C}))-B^{2}C\dot{A}(2(4\pi+\lambda)C\dot{B}
+B((8\pi+3\lambda)\right.\\\nonumber &\times& \left.\dot{C}-2\lambda
C'))+A^{2}(4(2\pi+\lambda)BCA'C'+(8\pi+3\lambda)C^{2}(A'B'\right.\\\nonumber
&-& \left.BA'')+
\lambda\dot{C}^{2}(-A'B'+BA''))+A^{3}(2q^{2}(4\pi+\lambda)B^{3}+q^{2}\lambda
B^{5}\right.\\\label{16}&+&\left.4(2\pi+\lambda)CB'C'-4(2\pi+\lambda)BCC'')\right].
\end{eqnarray}
The anisotropic parameter is defined as
\begin{equation}\label{15}
\triangle=p_{r}-p_{\theta}.
\end{equation}
To investigate the collapse and expansion of considered cylindrical
source, the expansion scalar is evaluated as
\begin{equation}\label{16}
\Theta=\frac{1}{A}\left(\frac{\dot{B}}{B}+\frac{\dot{C}}{C}\right),
\end{equation}
and an auxiliary solution of Eq.(\ref{9}) is
\begin{equation}\label{17}
A=\frac{\dot{C}}{\alpha C^{\gamma}},\quad B=\alpha C^{\gamma},
\end{equation}
where $\gamma$ and $\alpha>0$ are arbitrary constants. The above
solution leads to
\begin{equation}\label{18}
\Theta=\alpha(1+\gamma)C^{\gamma-1}.
\end{equation}
The positive values of $\Theta$ provide expansion and its negative
values correspond to collapse. The value of $\Theta$ depends on
$\alpha,~\gamma$ and $C$ in which $\alpha$ and $C$ are always
positive implying that we have collapse for $\gamma<-1$ and
expansion for $\gamma>-1$. We explore these cases one by one in the
following subsections.

\subsection{Collapse for $\gamma<-1$}

For collapsing solution, we find the unknown metric function $C$ in
the solution (\ref{17}) such that the collapse leads to the
formation of trapped surfaces. The mass function for the
cylindrically symmetric charged source is obtained as
\begin{equation}\label{19}
m(t,r)=\frac{1}{8}\left[1-\left(\frac{C'}{B}\right)^{2}+\left(\frac{\dot{C}}{A}\right)^{2}\right]+qC.
\end{equation}
Equation (\ref{17}) simplifies the mass expression as
\begin{equation}\label{20}
m(t,r)=\frac{1}{8}\left(1+\alpha^{2}C^{2\gamma}-\frac{C'^{2}}{\alpha^{2}C^{2\gamma}}\right)+qC.
\end{equation}
For trapped surface formation $m=\frac{1}{8}+qC$ \cite{8d}, which
yields
\begin{equation}\label{19}
C_{trap}=\left[\alpha^{2}(1-2\gamma)r+g(t)\right]^{\frac{1}{1-2\gamma}},
\end{equation}
where $g(t)$ is an integration function and the collapsing solution
becomes
\begin{eqnarray}\label{22}
A&=&\frac{1}{\alpha(1-2\gamma)}\dot{g}\left(\alpha^{2}(1-2\gamma)r+g(t)\right)^{\frac{\gamma}{1-2\gamma}},\\\label{23}
B&=&\alpha\left(\alpha^{2}(1-2\gamma)r+g(t)\right)^{\frac{\gamma}{1-2\gamma}},
\\\label{24}
C_{trap}&=&\left(\alpha^{2}(1-2\gamma)r+g(t)\right)^{\frac{1}{1-2\gamma}}.
\end{eqnarray}

For the sake of simplicity, we consider $g(t)$ as a linear function
of $t$, i.e., $g(t)=\frac{t}{\alpha^{2}}$ and obtain the following
expressions of density and pressures
\begin{eqnarray}\nonumber
\rho&=&\frac{\left(\frac{t}{\alpha^{2}}+r(1-2\gamma)\alpha^{2}\right)^{\frac{2+4\gamma}{2\gamma-1}}}
{8(1-2\gamma)^{2}(8\pi^{2}+6\pi\lambda+\lambda^{2})\alpha^{6})}
\left[q^{2}(8\pi+5\lambda)
\left(\frac{t}{\alpha^{2}}+r(1-2\gamma)\alpha^{2}\right)^{\frac{6\gamma}{1-2\gamma}}\right.\\\nonumber&-&\left.
\frac{1}{(t+r(1-2\gamma)\alpha^{4})^{2}}\left(2(1-2\gamma)^{2}\gamma(-4\pi+(-1+3\gamma)\lambda)\alpha^{12}
\right.\right.\\\nonumber&\times&\left.\left.
\left(\frac{t}{\alpha^{2}}+r(1-2\gamma)\alpha^{2}\right)^{\frac{2+2\gamma}{1-2\gamma}}\right)
+\frac{\gamma(-1+2\gamma)\lambda\alpha^{12}
\left(\frac{t}{\alpha^{2}}+r(1-2\gamma)\alpha^{2}\right)^{\frac{2+2\gamma}{1-2\gamma}}}
{(t+r(1-2\gamma)\alpha^{4})^{4}}\right.\\\nonumber&\times&\left.(1+t^{2}(1-2\gamma)^{2}-2rt(-1+2\gamma)^{3}\alpha^{4}
+r^{2}(1-2\gamma)^{4}\alpha^{8})-\frac{1}{(t+r(1-2\gamma)\alpha^{4})^{2}}\right.\\\nonumber&\times&\left.
\left((1-2\gamma)^{2}\gamma\lambda\alpha^{12}
\left(\frac{t}{\alpha^{2}}+r(1-2\gamma)\alpha^{2}\right)^{\frac{2+2\gamma}{1-2\gamma}}
(-1-2\alpha^{4}+\gamma(-2+4\alpha^{4}))\right)\right.\\\nonumber&+&\left.(1-2\gamma)^{2}\alpha^{3}
\left(\frac{t}{\alpha^{2}}+r(1-2\gamma)\alpha^{2}\right)^{\frac{\gamma}{1-2\gamma}}
\left\{-2q^{2}\lambda\alpha^{3}
\left(\frac{t}{\alpha^{2}}+r(1-2\gamma)\alpha^{2}\right)^{\frac{3\gamma}{1-2\gamma}}
\right.\right.\\\nonumber&+&\left.\left.q^{2}\lambda\alpha^{5}
\left(\frac{t}{\alpha^{2}}+r(1-2\gamma)\alpha^{2}\right)^{\frac{5\gamma}{1-2\gamma}}
+4\gamma(2\pi+\lambda)\alpha^{5}
\left(\frac{t}{\alpha^{2}}+r(1-2\gamma)\alpha^{2}\right)^{\frac{5\gamma}{1-2\gamma}}
\right.\right.\\\label{26}&-&\left.\left.\frac{1}{(t+r(1-2\gamma)\alpha^{4})^{2}}8\gamma(2\pi+\lambda)\alpha^{9}
\left(\frac{t}{\alpha^{2}}+r(1-2\gamma)\alpha^{2}\right)^{\frac{2+\gamma}{1-2\gamma}}\right\}\right],\\\nonumber
p_{r}&=&\frac{\left(\frac{t}{\alpha^{2}}+r(1-2\gamma)\alpha^{2}\right)^{\frac{2+4\gamma}{2\gamma-1}}}
{8(1-2\gamma)^{2}(2\pi^{2}+\lambda)(4\pi^{2}+\lambda)\alpha^{6})}\left[-q^{2}\lambda
\left(\frac{t}{\alpha^{2}}+r(1-2\gamma)\alpha^{2}\right)^{\frac{6\gamma}{1-2\gamma}}
\right.\\\nonumber&+&\left.\frac{1}{(t+r(1-2\gamma)\alpha^{4})^{2}}\left(2(1-2\gamma)^{2}\gamma
(-8\pi+3(-1+\gamma)\lambda)\alpha^{12}\right.\right.\\\nonumber&\times&\left.\left.
\left(\frac{t}{\alpha^{2}}+r(1-2\gamma)\alpha^{2}\right)^{\frac{2+2\gamma}{1-2\gamma}}\right)+
(q-2q\gamma)^{2}\alpha^{6}\left(\frac{t}{\alpha^{2}}+r(1-2\gamma)\alpha^{2}\right)^{\frac{4\gamma}{1-2\gamma}}
\right.\\\nonumber&\times&\left.
(2\lambda+(8\pi+3\lambda)\alpha^{2}\left(\frac{t}{\alpha^{2}}+r(1-2\gamma)\alpha^{2}\right)^{\frac{2\gamma}{1-2\gamma}})
-(-8\pi(-1+2\gamma)\right.\\\nonumber&\times&\left.
(t+r(1-2\gamma)\alpha^{4})^{2}+\lambda(1+t^{2}(5-12\gamma+4\gamma^{2})
-2rt(1-2\gamma)^{2}(-5+2\gamma)\alpha^{4}\right.\\\nonumber
&+&\left.r^{2}(-5+2\gamma)(-1+2\gamma)^{3}\alpha^{3}))
((t+r(1-2\gamma)\alpha^{4})^{4}\gamma(-1+2\gamma)\alpha^{12}\right.\\\nonumber&\times&\left.
\left(\frac{t}{\alpha^{2}}+r(1-2\gamma)\alpha^{2}\right)^{\frac{2+2\gamma}{1-2\gamma}})
-\frac{1}{(t+r(1-2\gamma)\alpha^{4})^{2}}(1-2\gamma)^{2}\gamma\alpha^{12}\right.\\\nonumber&\times&\left.
\left(\frac{t}{\alpha^{2}}+r(1-2\gamma)\alpha^{2}\right)^{\frac{2+2\gamma}{1-2\gamma}}
(16\pi(-1+2\gamma)\alpha^{4}\right.\\\label{27}&+&\left.\lambda(1-6\alpha^{4}+2\gamma(1+6\alpha^{4})))\right],\\\nonumber
p_{\theta}&=&\frac{\left(\frac{t}{\alpha^{2}}+r(1-2\gamma)\alpha^{2}\right)^{\frac{2+4\gamma}{-1+2\gamma}}}
{8(1-2\gamma)^{2}(2\pi+\lambda)(4\pi+\lambda)\alpha^{6}}\left[-q^{2}\lambda
\left(\frac{t}{\alpha^{2}}+r(1-2\gamma)\alpha^{2}\right)^{\frac{6\gamma}{1-2\gamma}}\right.\\\nonumber
&-&\left.\frac{2(1-2\gamma)^{2}\gamma(4\pi(-1+3\gamma)+3(-1+\gamma)\lambda)\alpha^{12}
\left(\frac{t}{\alpha^{2}}+r(1-2\gamma)\alpha^{2}\right)^{\frac{2+2\gamma}{1-2\gamma}}}
{(t+r(1-2\gamma)\alpha^{4})^{2}}\right.\\\nonumber&+&\left.\frac{(1-2\gamma)^{2}\gamma\alpha^{12}
\left(\frac{t}{\alpha^{2}}+r(1-2\gamma)\alpha^{2}\right)^{\frac{2+2\gamma}{1-2\gamma}}}
{(t+r(1-2\gamma)\alpha^{4})^{2}}(8\pi\gamma+(-1+2\gamma)\lambda(1+2\alpha^{4}))
\right.\\\nonumber&+&\left.\frac{(1-2\gamma)^{3}\gamma\alpha^{12}
\left(\frac{t}{\alpha^{2}}+r(1-2\gamma)\alpha^{2}\right)^{\frac{2+2\gamma}{1-2\gamma}}}
{(t+r(1-2\gamma)\alpha^{4})^{2}}\left(\lambda-\frac{8\pi+3\lambda}{(1-2\gamma)^{2}(t+r(1-2\gamma)\alpha^{4})^{2}}
\right.\right.\\\nonumber&-&\left.\left.q^{2}(1-2\gamma)^{2}\alpha^{6}\right)
\left(\frac{t}{\alpha^{2}}+r(1-2\gamma)\alpha^{2}\right)^{\frac{4\gamma}{1-2\gamma}}
\left(8\pi+\lambda\left(2+\alpha^{2}\right.\right.\right.\\\label{28}&\times&\left.\left.\left.
\left(\frac{t}{\alpha^{2}}+r(1-2\gamma)\alpha^{2}\right)^{\frac{2\gamma}{1-2\gamma}}\right)\right)
\right],\\\nonumber
p_{z}&=&\frac{\alpha^{6}\left(\frac{t}{\alpha^{2}}+r(1-2\gamma)\alpha^{2}\right)^{\frac{2+2\gamma}{-1+2\gamma}}}
{8(1-2\gamma)^{2}(2\pi+\lambda)(4\pi+\lambda)}\left[8\pi(1-2\gamma)^{2}
\left(\frac{-2\gamma
\left(\frac{t}{\alpha^{2}}+r(1-2\gamma)\alpha^{2}\right)^{\frac{2}{1-2\gamma}}}
{(t+r(1-2\gamma)\alpha^{4})^{2}}\right.\right.\\\nonumber&-&\left.\left.\frac{q^{2}
\left(\frac{t}{\alpha^{2}}+r(1-2\gamma)\alpha^{2}\right)^{\frac{2\gamma}{1-2\gamma}}}
{\alpha^{6}}\right)+\lambda\left(\frac{-q^{2}
\left(\frac{t}{\alpha^{2}}+r(1-2\gamma)\alpha^{2}\right)^{\frac{2\gamma}{1-2\gamma}}}
{\alpha^{12}}\right.\right.\\\nonumber&\times&\left.\left.\left(2(1-2\gamma)^{2}\alpha^{6}+
\left(\frac{t}{\alpha^{2}}+r(1-2\gamma)\alpha^{2}\right)^{\frac{2\gamma}{1-2\gamma}}+
(1-2\gamma)^{2}\alpha^{6}\right.\right.\right.\\\nonumber&\times&\left.\left.\left.
\left(\frac{t}{\alpha^{2}}+r(1-2\gamma)\alpha^{2}\right)^{\frac{2\gamma}{1-2\gamma}}\right)+
(-1+(4rt(1-2\gamma)^{2}\alpha^{4}+2r^{2}(-1+2\gamma)^{3}\alpha^{8})
\right.\right.\\\nonumber&\times&\left.\left.
(-3+\gamma-\alpha^{4}+2\gamma\alpha^{4}+2t^{2}(3+\alpha^{4}+
\gamma^{2}(2+4\alpha^{4})-\gamma(7+4\alpha^{4}))))\right.\right.\\\label{29}&\times&\left.\left.
\left((t+r(1-2\gamma)\alpha^{4})^{4}\gamma(-1+2\gamma)
\left(\frac{t}{\alpha^{2}}+r(1-2\gamma)\alpha^{2}\right)^{\frac{2}{1-2\gamma}}\right)^{-1}\right)\right].
\end{eqnarray}
For the solution (\ref{22})-(\ref{24}), the anisotropic parameter
and mass function become
\begin{eqnarray}\nonumber
\triangle&=&\frac{\alpha^{6}\left(\frac{t}{\alpha^{2}}+r(1-2\gamma)\alpha^{2}\right)^{\frac{2+2\gamma}{-1+2\gamma}}}
{8(1-2\gamma)(4\pi+\lambda)}\left[-q^{2}(-1+\gamma)\alpha^{-6}
\left(\frac{t}{\alpha^{2}}+r(1-2\gamma)\alpha^{2}\right)^{\frac{2\gamma}{1-2\gamma}}\right.\\\nonumber&\times&\left.
\left(1+\alpha^{2}\left(\frac{t}{\alpha^{2}}+r(1-2\gamma)\alpha^{2}\right)^{\frac{2\gamma}{-1+2\gamma}}\right)
+\left(1-4rt(1-2\gamma)^{2}\alpha^{4}(1-\alpha^{4}\right.\right.\\\nonumber&-&\left.\left.\gamma(-1+2\alpha^{4}))+2r^{2}(-1+2\gamma)^{3}\alpha^{8}
(1-\alpha^{8}-\gamma(-1+2\alpha^{4}))\right.\right.\\\label{29a}&+&\left.\left.\frac{2t^{2}(-1+\alpha^{4}+
\gamma(3-4\alpha^{4})+\gamma^{2}(-2+4\alpha^{4}))}{(t+r(1-2\gamma)\alpha^{4})^{4}\gamma
\left(\frac{t}{\alpha^{2}}+r(1-2\gamma)\alpha^{2}\right)^{\frac{2}{-1+2\gamma}}}\right)\right],
\\\label{29}
m&=&\frac{1}{8}+q\left(\frac{t}{\alpha^{2}}+r(1-2\gamma)\alpha^{2}\right)^{\frac{1}{1-2\gamma}}.
\end{eqnarray}
\begin{figure}
\center\epsfig{file=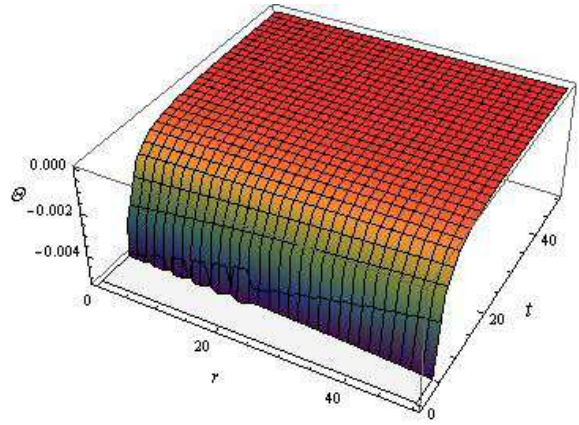,width=0.55\linewidth}\caption{Plot of
$\Theta$ versus $r$ and $t$ for $\gamma=-1.5$, $\alpha=0.1$.}
\end{figure}
\begin{figure}
\epsfig{file=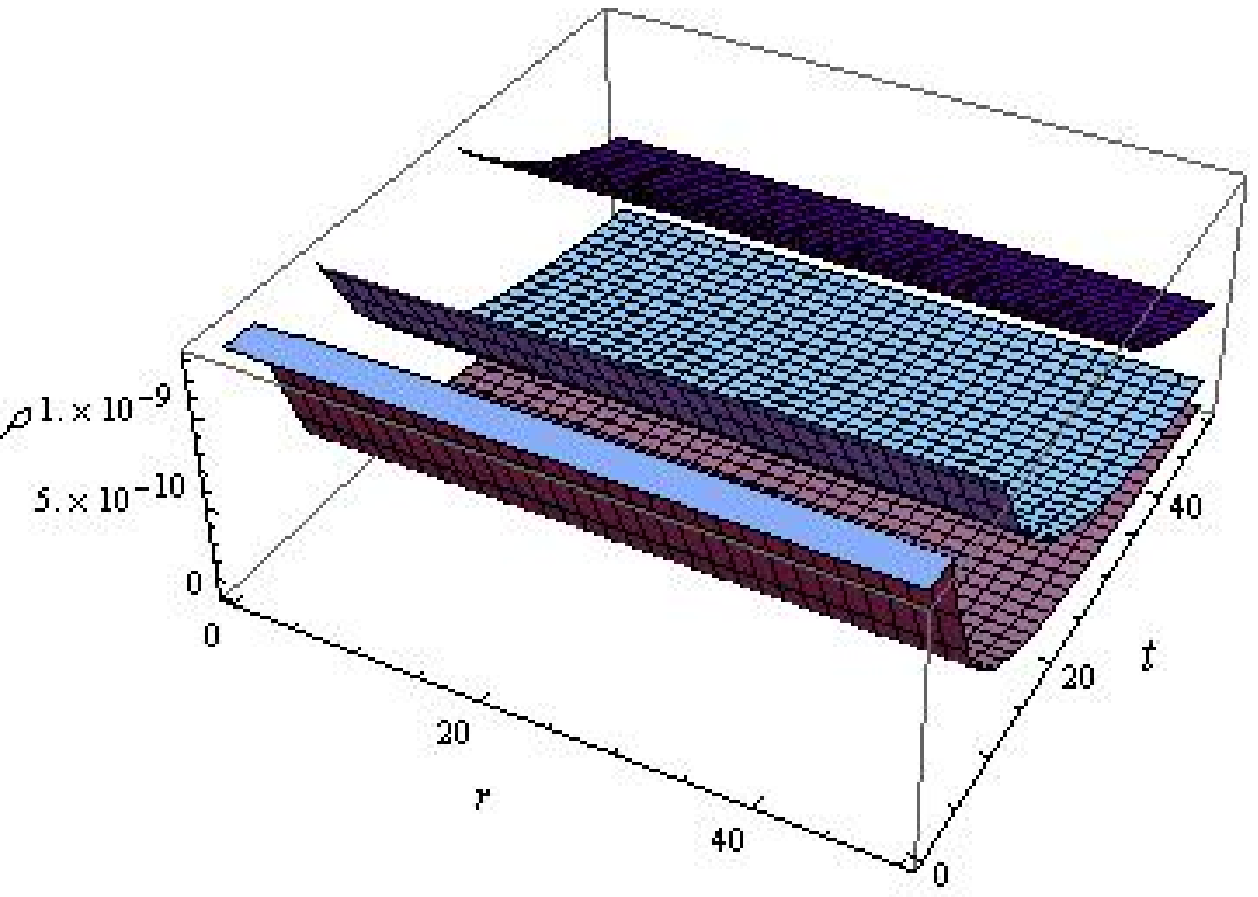,width=0.55\linewidth}\epsfig{file=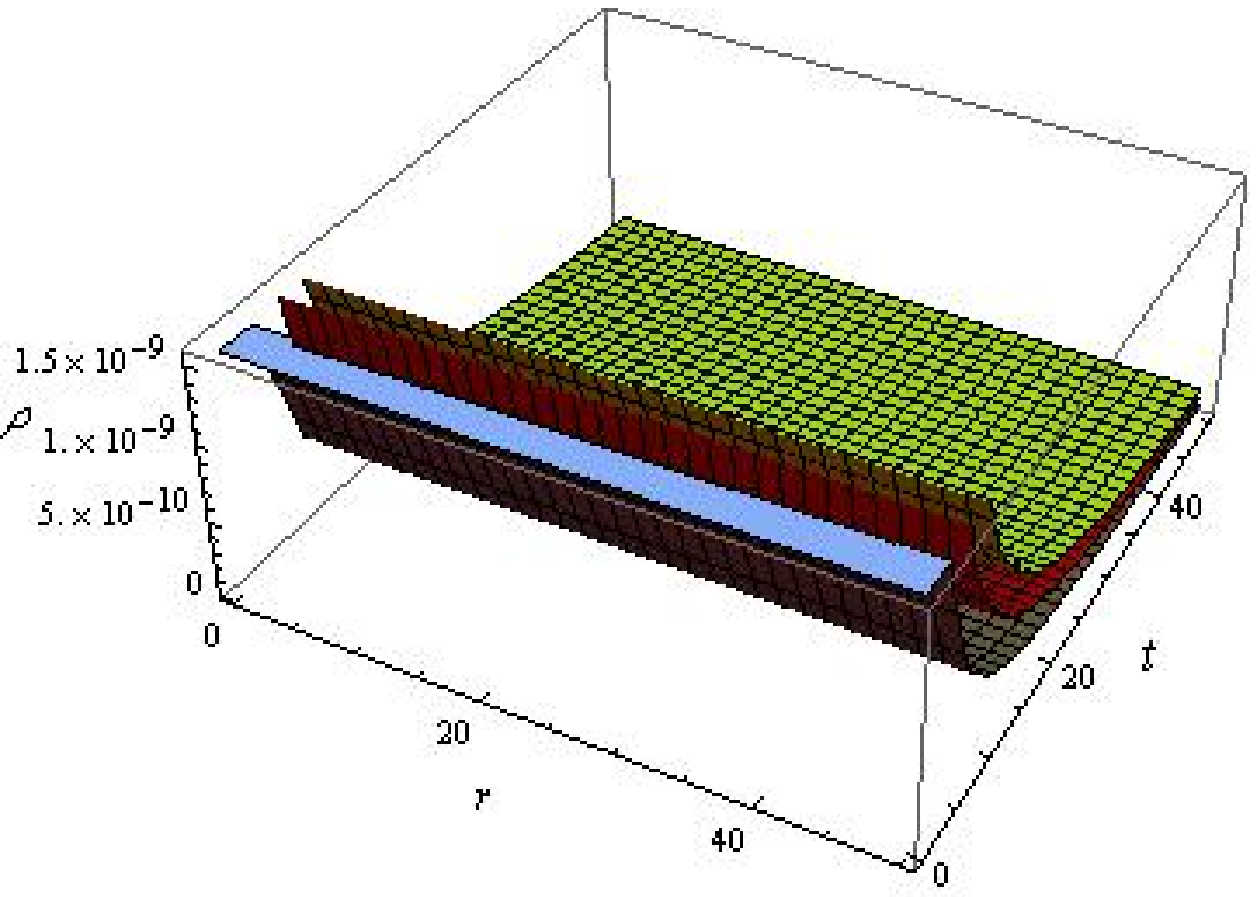,width=0.55\linewidth}
\caption{Plots of $\rho$ versus $r$ and $t$ for $\gamma=-1.5$,
$\alpha=0.1$. The left graph is for $q=0$ (pink), $q=0.00005$
(blue), $q=0.0001$ (purple) with $\lambda=-0.1$ and the right graph
is for $\lambda=-0.1$ (brown), $\lambda=-0.2$ (red), $\lambda=-0.3$
(yellow) with $q=0.0001$.}
\end{figure}
\begin{figure}
\epsfig{file=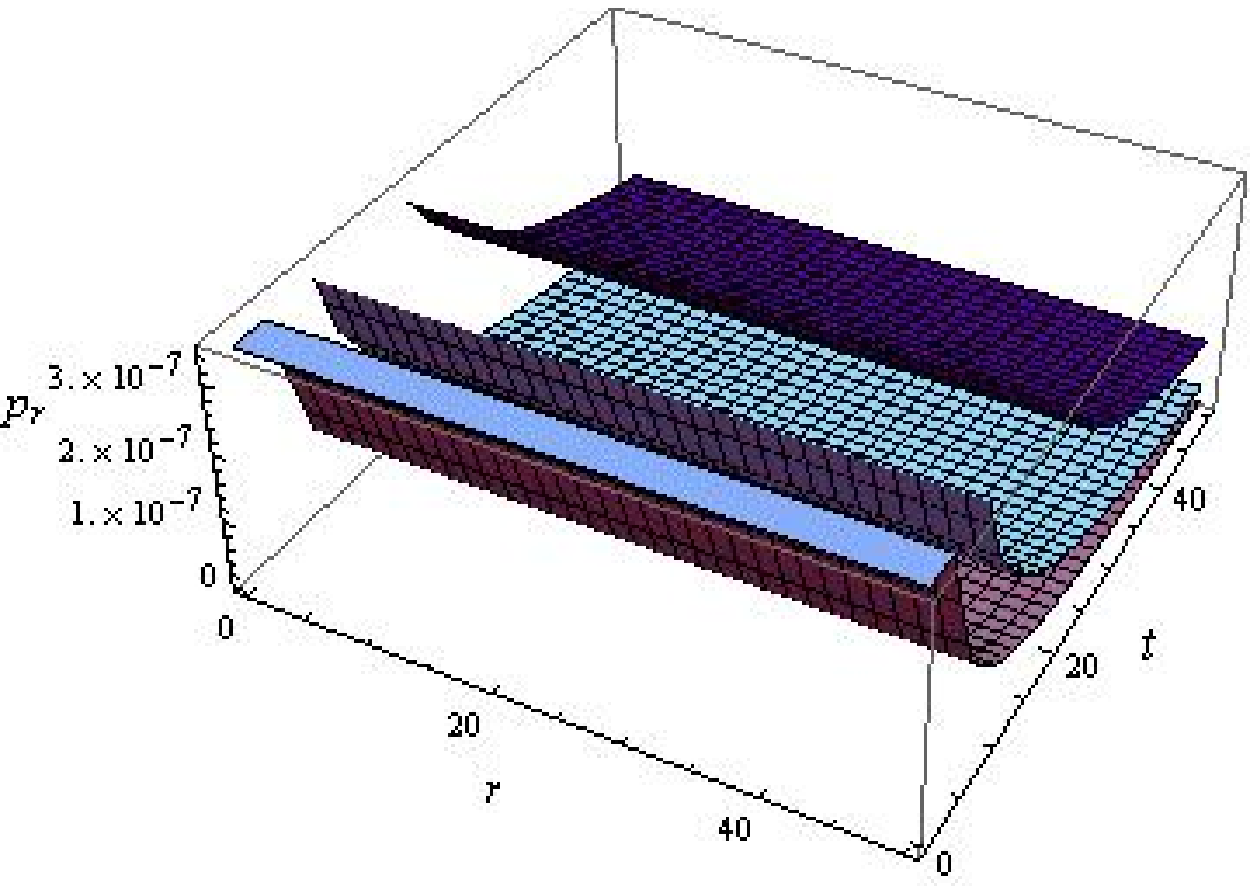,width=0.55\linewidth}\epsfig{file=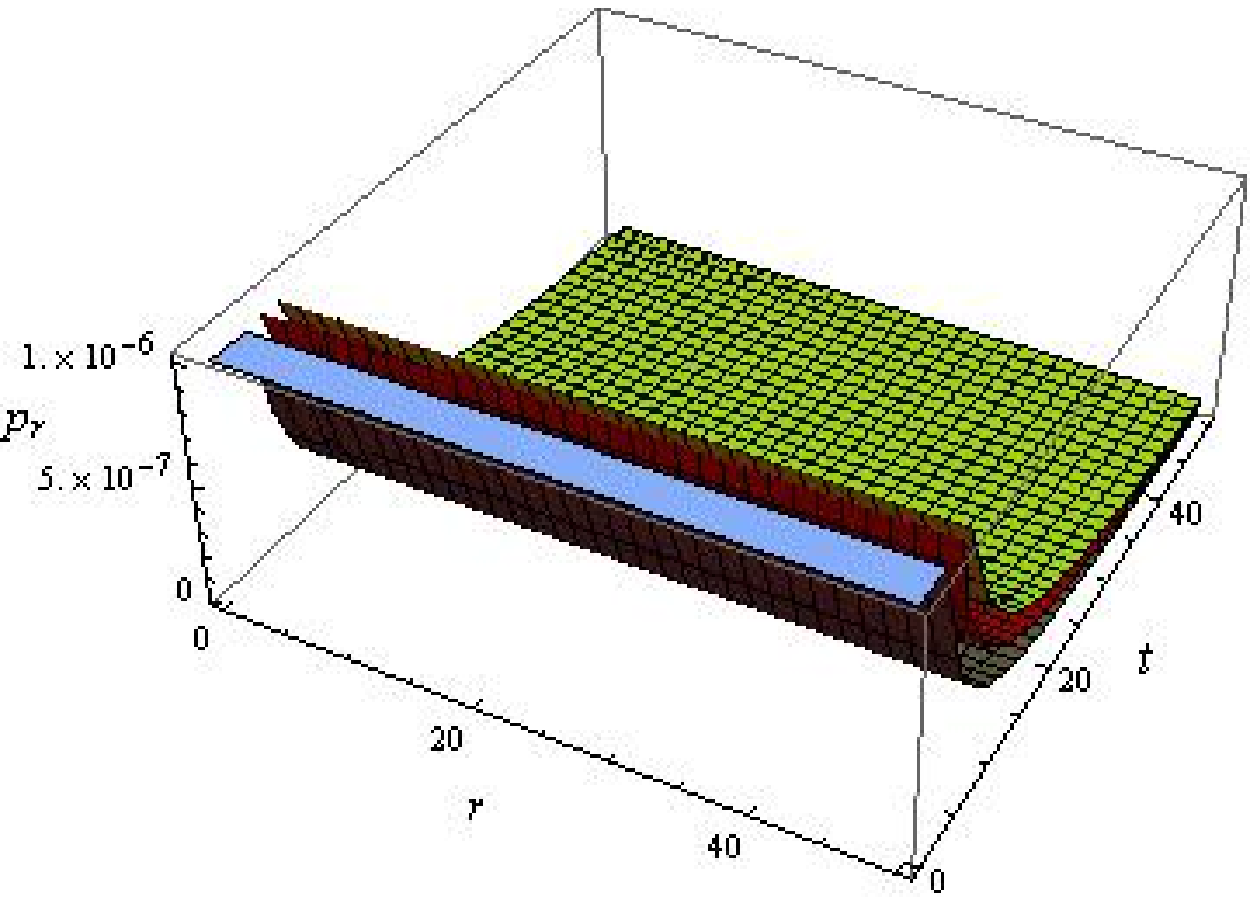,width=0.55\linewidth}
\caption{Plots of $p_{r}$ versus $r$ and $t$ for $\gamma=-1.5$,
$\alpha=0.1$. The left graph is for $q=0$ (pink), $q=0.01$ (blue),
$q=0.02$ (purple) with $\lambda=-0.1$ while the right graph is for
$\lambda=-0.1$ (brown), $\lambda=-0.2$ (red), $\lambda=-0.3$
(yellow) with $q=0.01$.}
\end{figure}
\begin{figure}
\epsfig{file=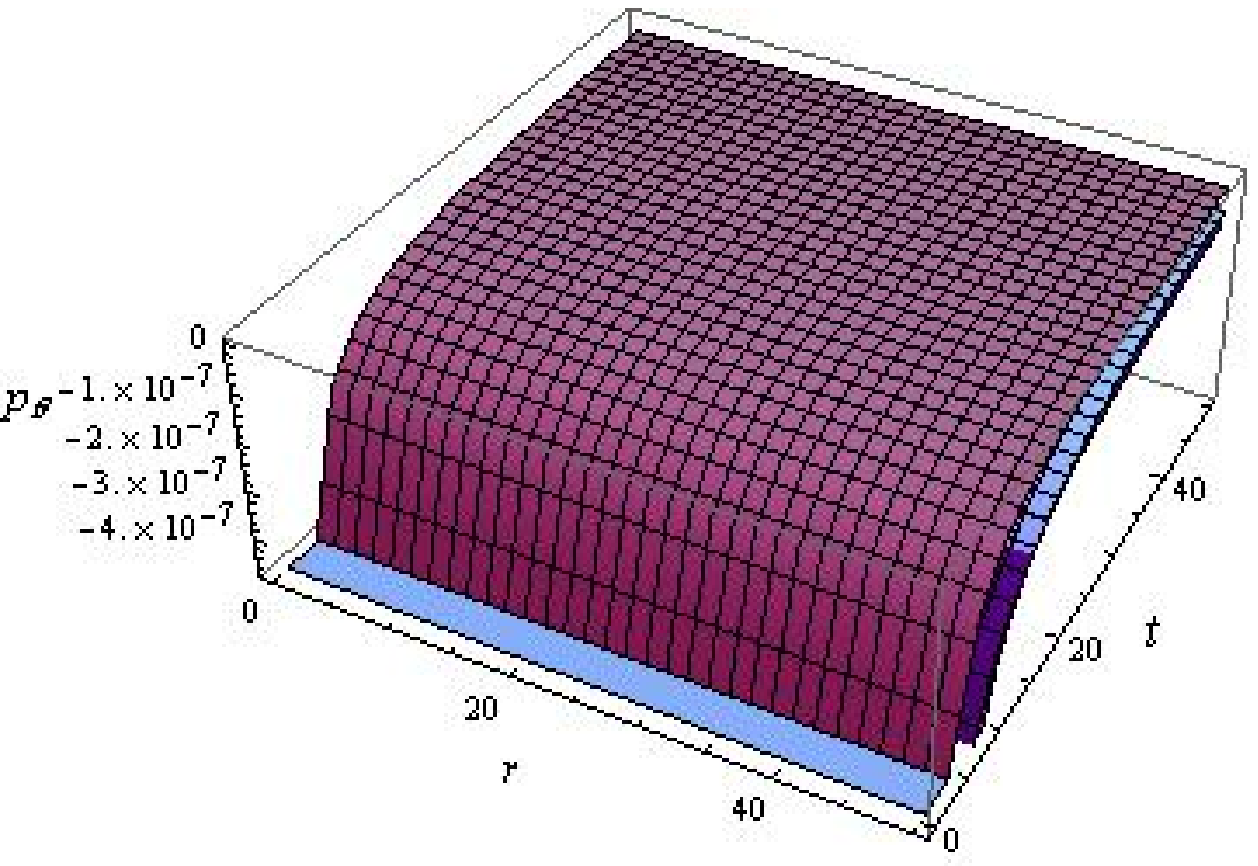,width=0.55\linewidth}\epsfig{file=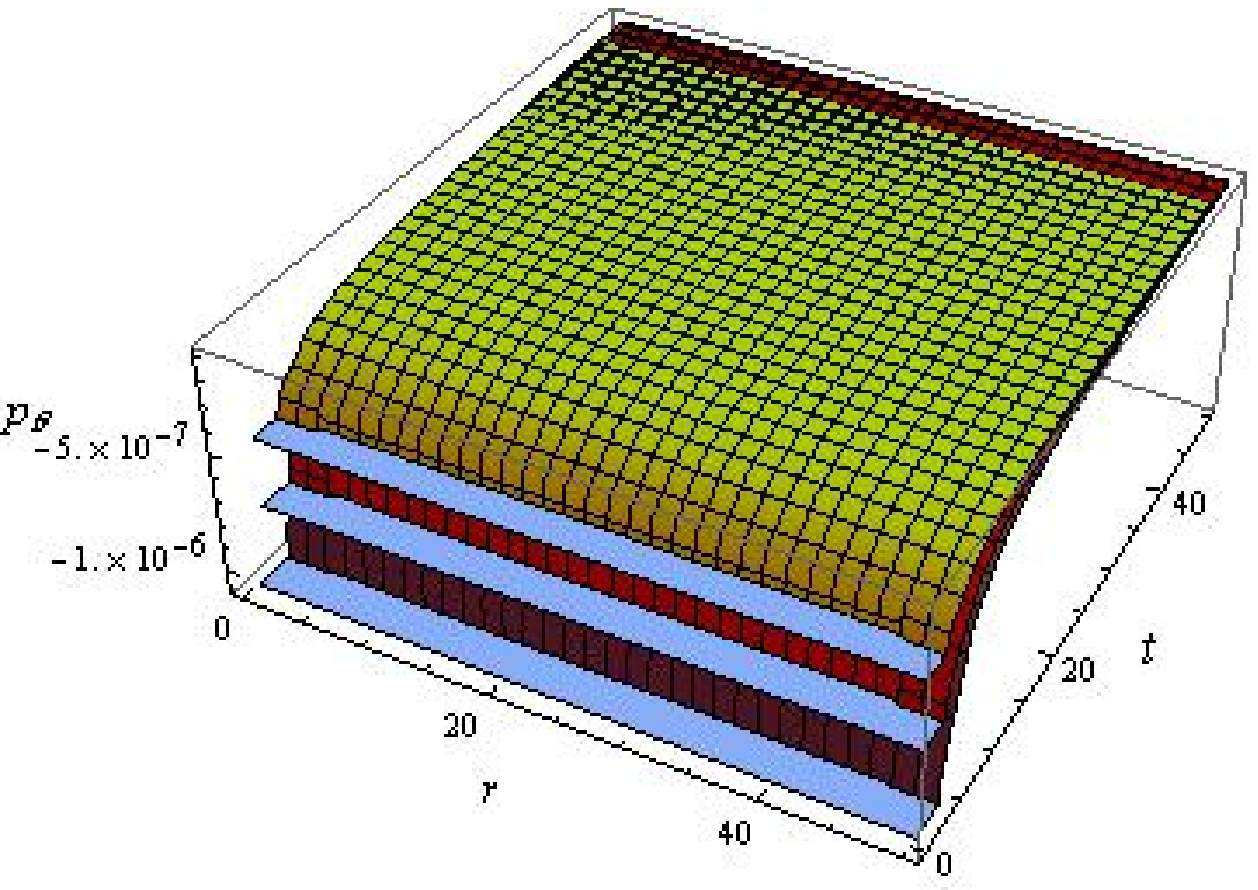,width=0.55\linewidth}
\caption{Plots of $p_{\theta}$ versus $r$ and $t$ for $\gamma=-1.5$,
$\alpha=0.1$. The left graph is for $q=0$ (pink), $q=0.005$ (blue),
$q=0.01$ (purple) with $\lambda=-0.1$ and the right graph is for
$\lambda=-0.1$ (brown), $\lambda=-0.15$ (red), $\lambda=-0.2$
(yellow) with $q=0.01$.}
\end{figure}
\begin{figure}
\epsfig{file=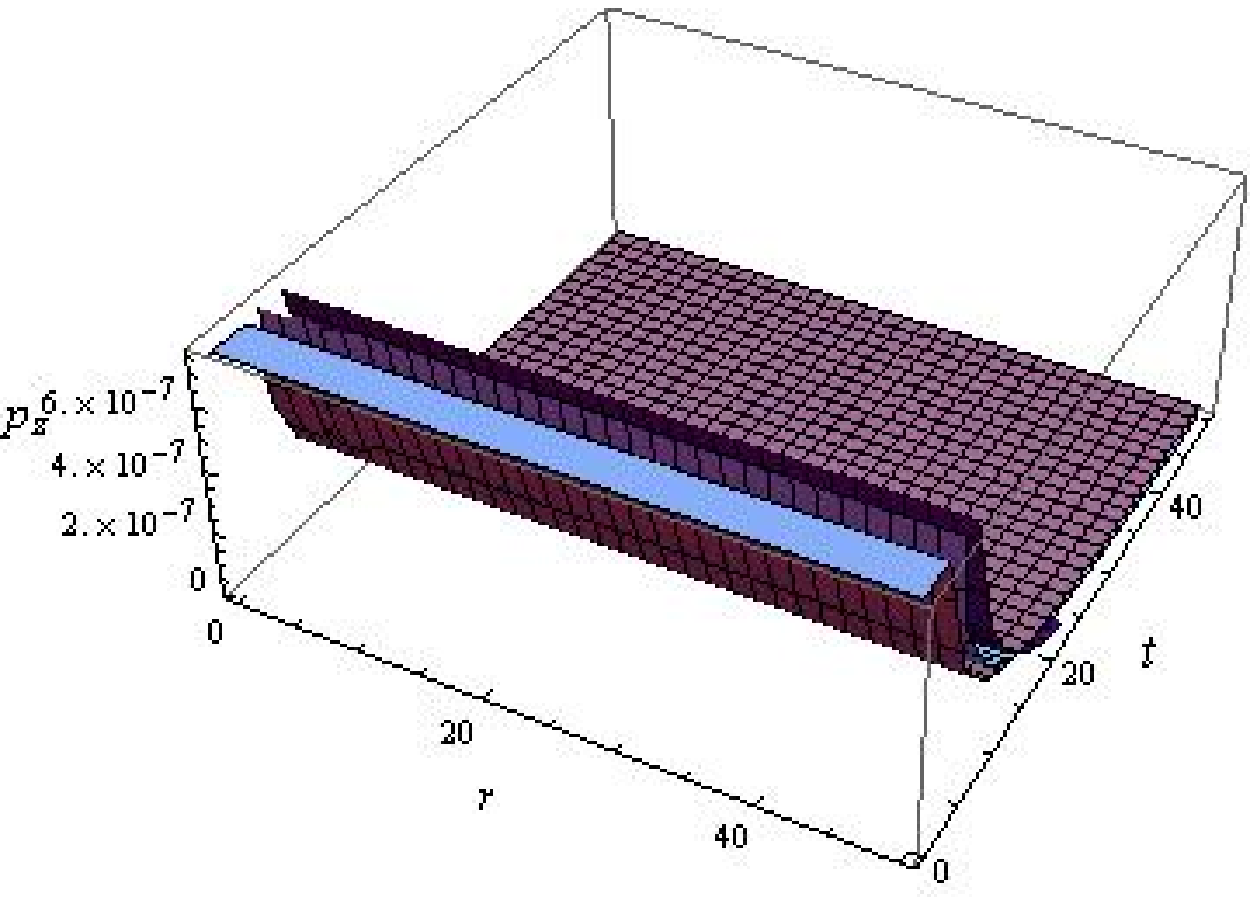,width=0.55\linewidth}\epsfig{file=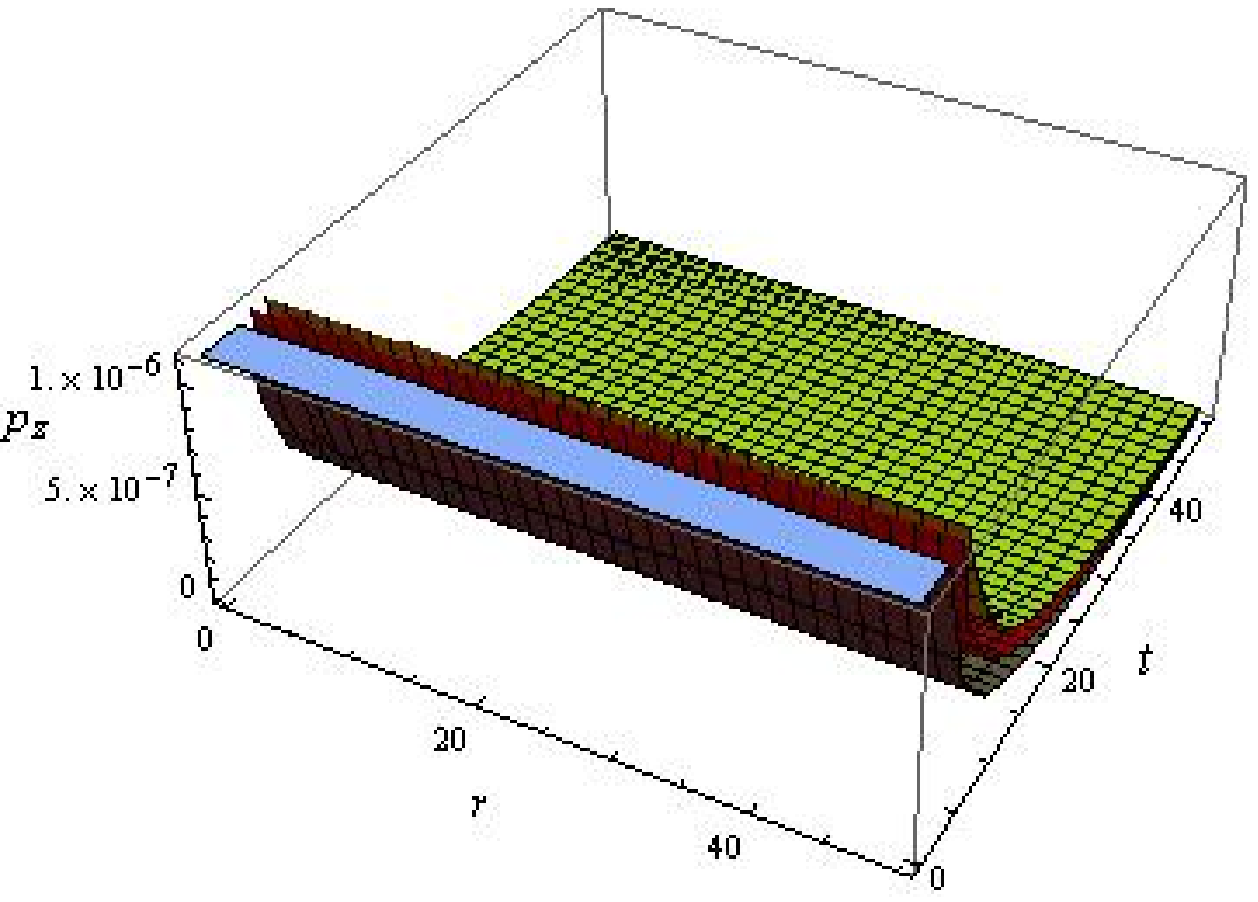,width=0.55\linewidth}
\caption{Plots of $p_{z}$ versus $r$ and $t$ for $\gamma=-1.5$,
$\alpha=0.1$. The left graph is for $q=0$ (pink), $q=0.01$ (blue),
$q=0.02$ (purple) with $\lambda=-0.1$ and the right graph is for
$\lambda=-0.1$ (brown), $\lambda=-0.2$ (red), $\lambda=-0.3$
(yellow) with $q=0.01$.}
\end{figure}
\begin{figure}
\epsfig{file=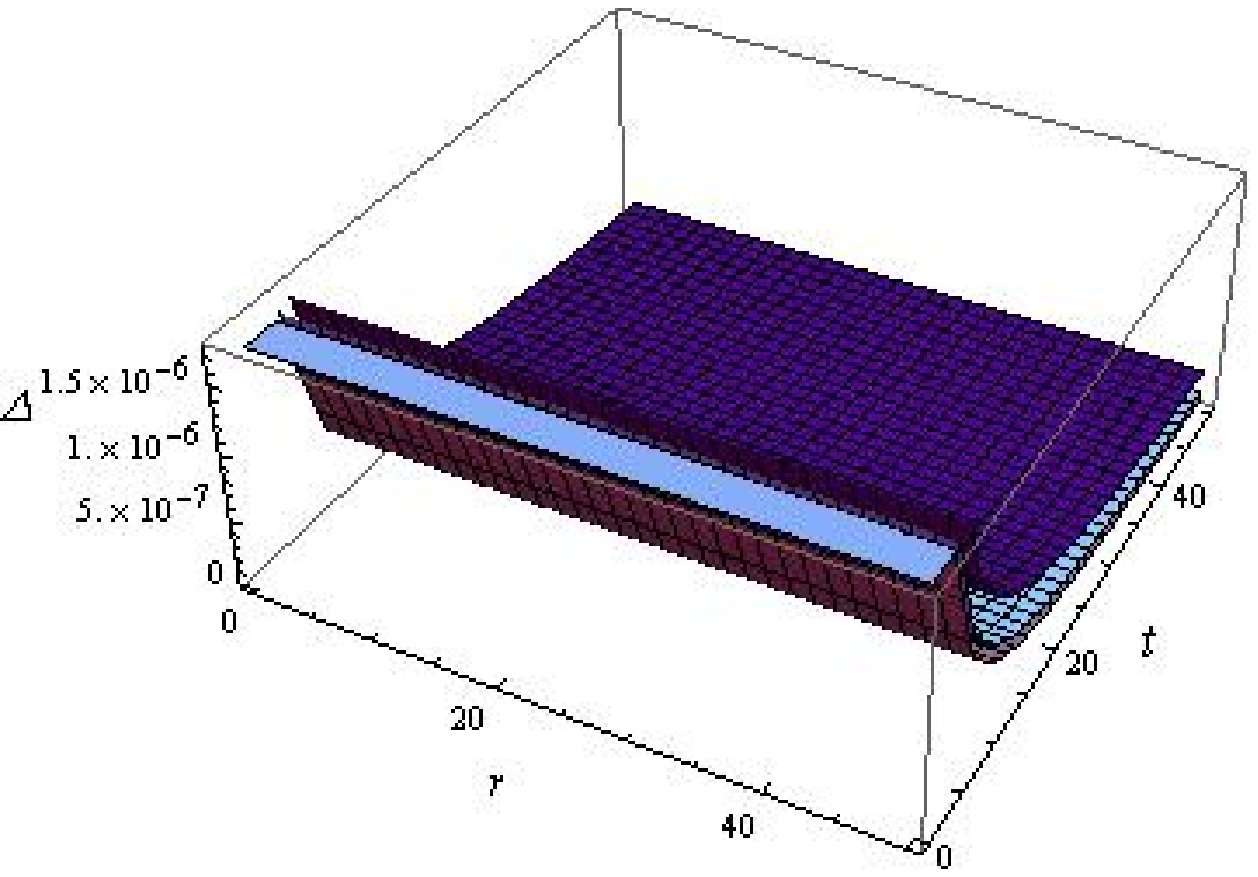,width=0.55\linewidth}\epsfig{file=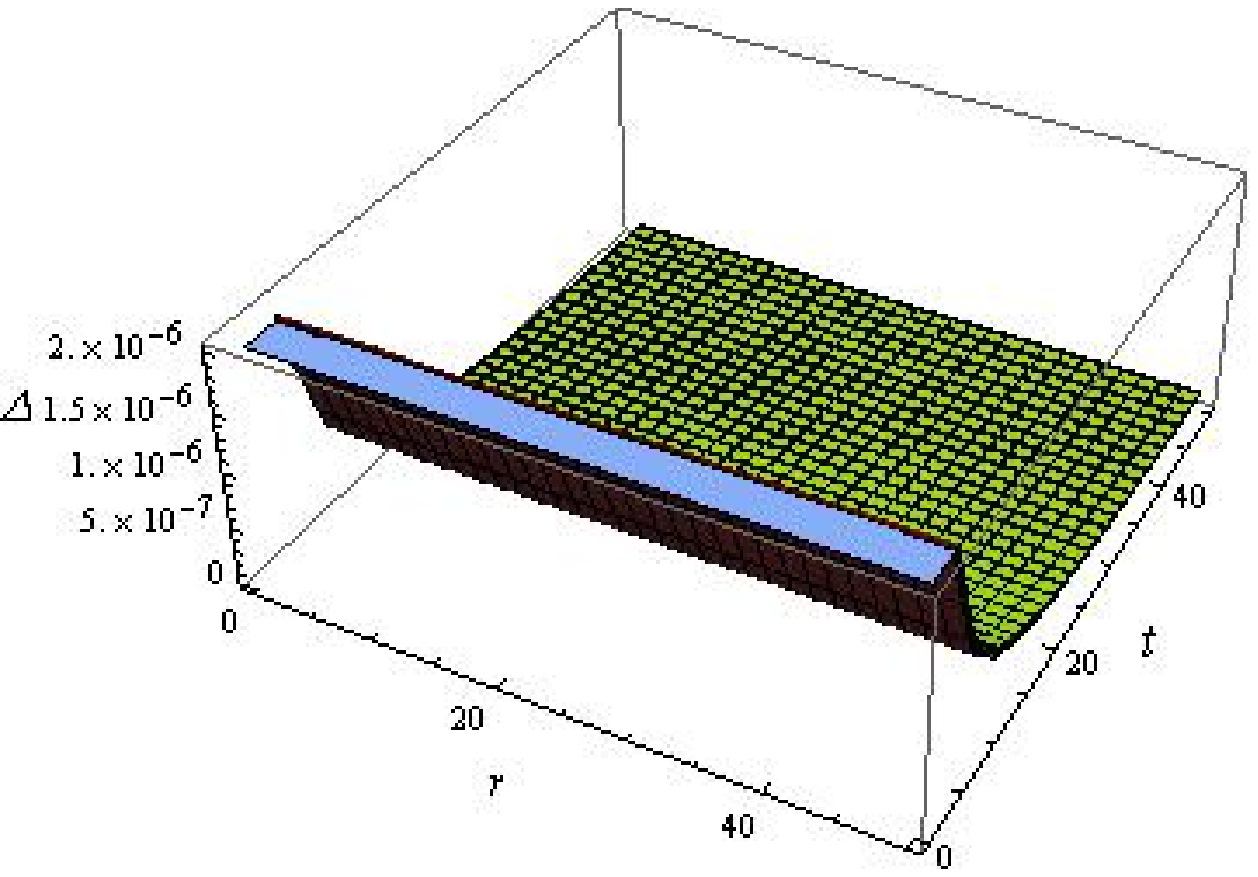,width=0.55\linewidth}\caption{Plots
of $\triangle$ versus $r$ and $t$ for $\gamma=-1.5$, $\alpha=0.1$.
The left graph is for $q=0$ (pink), $q=0.01$ (blue), $q=0.02$
(purple) with $\lambda=-0.1$ and the right graph is for
$\lambda=-0.1$ (brown), $\lambda=-0.2$ (red), $\lambda=-0.3$
(yellow) with $q=0.01$.}
\end{figure}

For the collapsing case, the graphical representation of different
parameters is given in Figures \textbf{1}-\textbf{7}. We observe
that the quantities are changing with respect to temporal coordinate
while no change is observed with respect to radial coordinate. The
change in different quantities with respect to time is given in
Table \textbf{1} and the effects of charge as well as model
parameter $\lambda$ are summarized in Table \textbf{2}.\\\\
Table \textbf{1}: Change in parameters with respect to $t$ for the
collapse solution
\begin{table}[bht]
\centering
\begin{small}
\begin{tabular}{|c|c|c|c|c|c|c|}
\hline\textbf{Parameter}&$\rho$&$p_{r}$&$p_{\theta}$&$p_{z}$&$\triangle$&$m$\\
\hline\textbf{As $t$ increases}&decreases&decreases&increases&decreases&decreases&increases\\
\hline
\end{tabular}
\end{small}
\end{table}\\
Table \textbf{2}: Effects of $q$ and $\lambda$ for the collapse
solution
\begin{table}[bht]
\centering
\begin{small}
\begin{tabular}{|c|c|c|c|c|c|c|}
\hline\textbf{Parameter}&$\rho$&$p_{r}$&$p_{\theta}$&$p_{z}$&$\triangle$&$m$\\
\hline\textbf{As $q$ increases}&increases&increases&decreases&decreases&increases&increases\\
\hline\textbf{As $\lambda$ decreases}&increases&increases&increases&increases&increases&no change\\
\hline
\end{tabular}
\end{small}
\end{table}
\begin{figure}
\center\epsfig{file=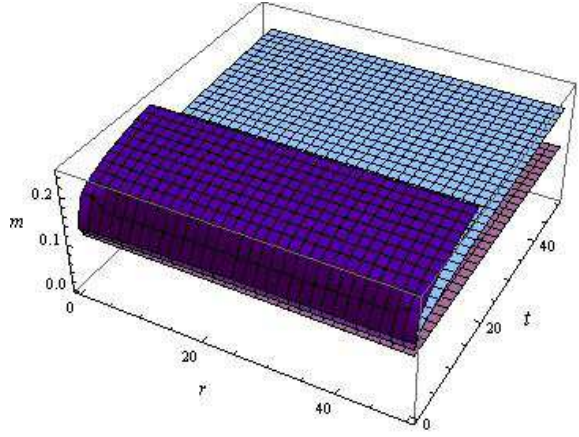,width=0.55\linewidth}\caption{Plot of $m$
versus $r$ and $t$ for $\gamma=-1.5$, $\alpha=0.1$, $q=0$ (pink),
$q=0.01$ (blue), $q=0.02$ (purple).}
\end{figure}

To observe physical viability of our solution, we plot the null
(NEC), weak (WEC), strong (SEC) and dominant (DEC) energy conditions
for the curvature-matter coupled gravity \cite{Td1}
\begin{itemize}
\item{NEC: $\rho+p_{r}-\mathcal{A}\geq0,~
\rho+p_{\theta}-\mathcal{A}\geq0,~\rho+p_{z}-\mathcal{A}\geq0$,}
\item{WEC: $\rho-\mathcal{A}\geq0,
~\rho+p_{r}-\mathcal{A}\geq0,~\rho+p_{\theta}-\mathcal{A}\geq0,~\rho+p_{z}-\mathcal{A}\geq0,$}
\item{SEC: $\rho+p_{r}+p_{\theta}+p_{z}-\mathcal{A}\geq0$.}
\item{DEC: $\rho-p_{r}-\mathcal{A}\geq0,~\rho+p_{\theta}-\mathcal{A}\geq0,
~\rho-p_{z}-\mathcal{A}\geq0$,}
\end{itemize}
The term $\mathcal{A}=(V^{\beta}V^{\alpha}_{;\beta})_{;\alpha}$ is
due to non-geodesic motion of massive particles. We evaluate
$\mathcal{A}$ as
\begin{equation}
\mathcal{A}=\frac{1}{B^{2}}\left[\frac{A''}{A}+\frac{A'}{A}\left(-
\frac{B'}{B}+\frac{C'}{C}+\frac{A'}{A}\right)\right]+\frac{\dot{A}^{2}}{A^{4}}.
\end{equation}
For the collapse solution, we have
\begin{equation}
\mathcal{A}=\frac{4\gamma^{2}\alpha^{6}
\left(\frac{t}{\alpha^{2}}+r(1-2\gamma)\alpha^{2}\right)^{\frac{2\gamma}{-1+2\gamma}}}
{(t+r(1-2\gamma)\alpha^{4})^{2}}.
\end{equation}
All the energy conditions defined above are plotted in Figures
\textbf{8}-\textbf{11}, the repeated expressions are shown only
once. From these plots, it can be easily seen that all the energy
conditions are satisfied for the considered values of free
parameters of the collapse solution.
\begin{figure}
\epsfig{file=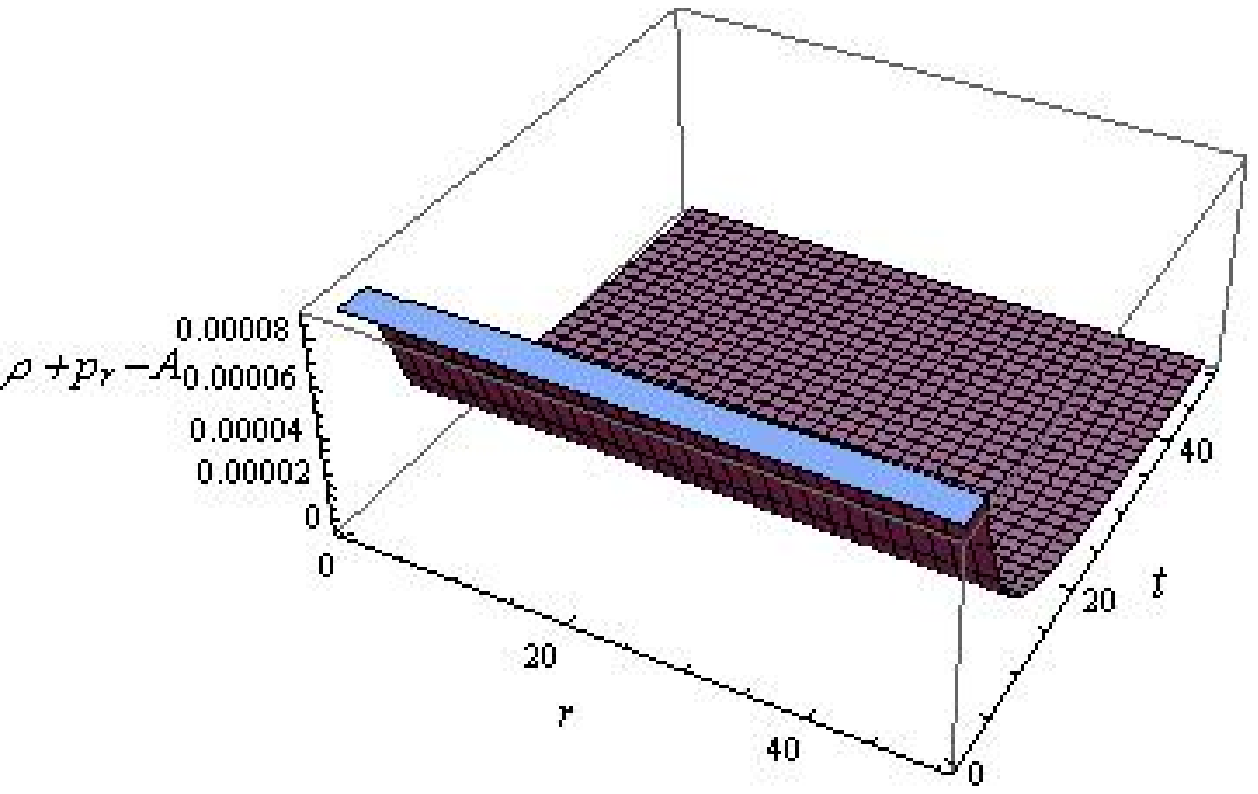,width=0.55\linewidth}\epsfig{file=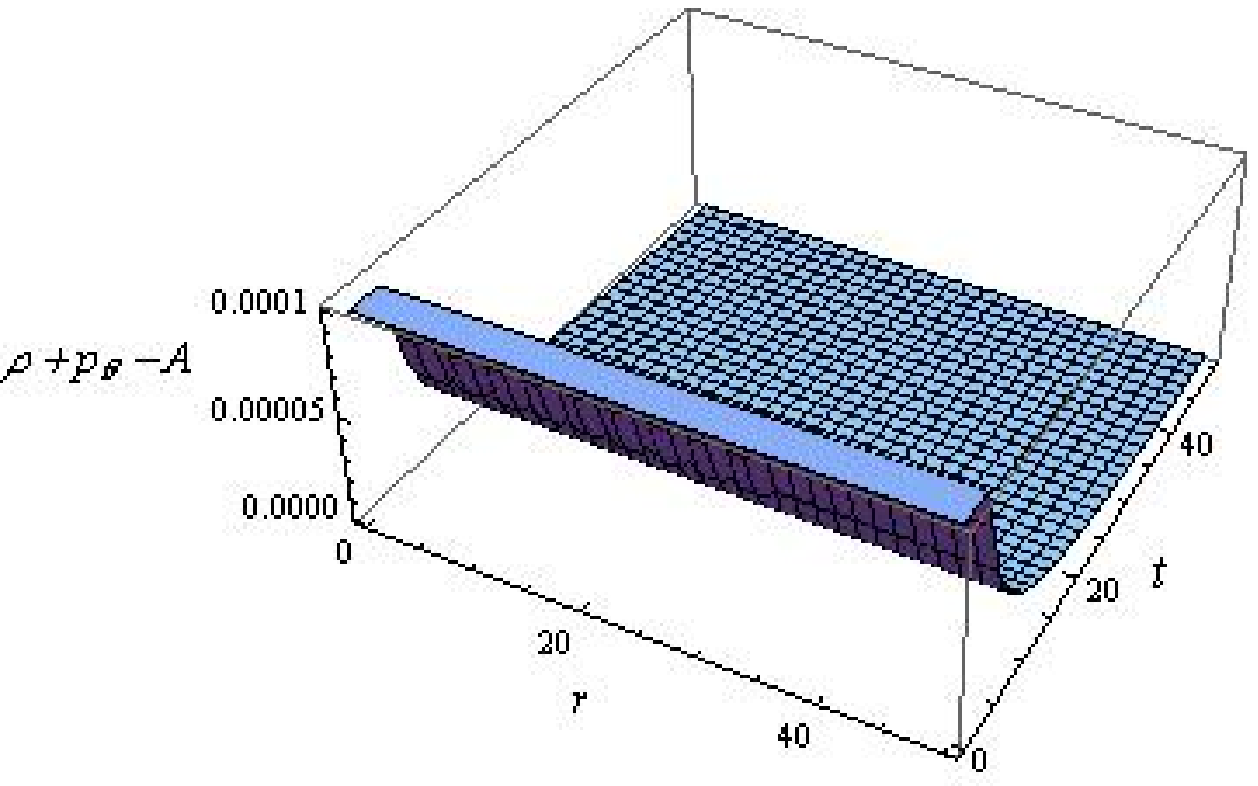,width=0.55\linewidth}
\center\epsfig{file=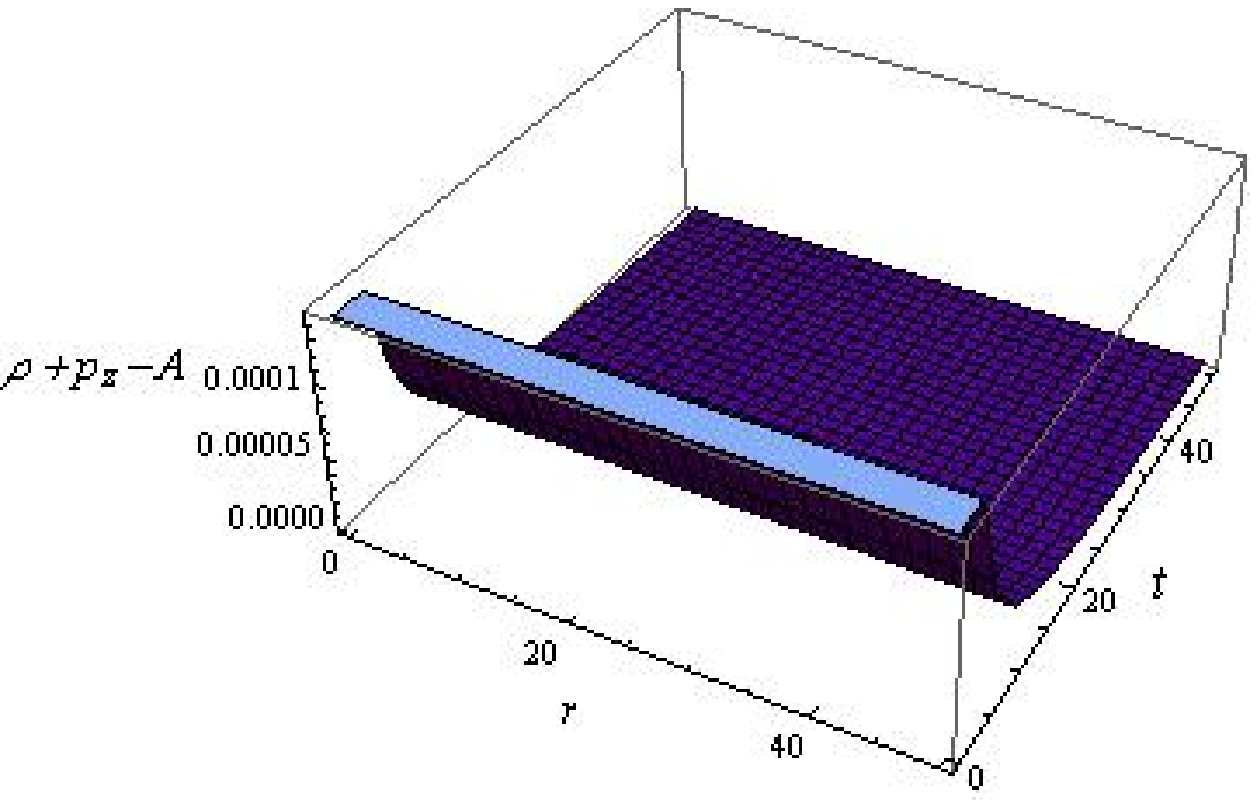,width=0.55\linewidth} \caption{Plots
for NEC for $\gamma=-1.5$, $\alpha=0.1$, $q=0.01$ and
$\lambda=-0.1$.}
\end{figure}
\begin{figure}
\center\epsfig{file=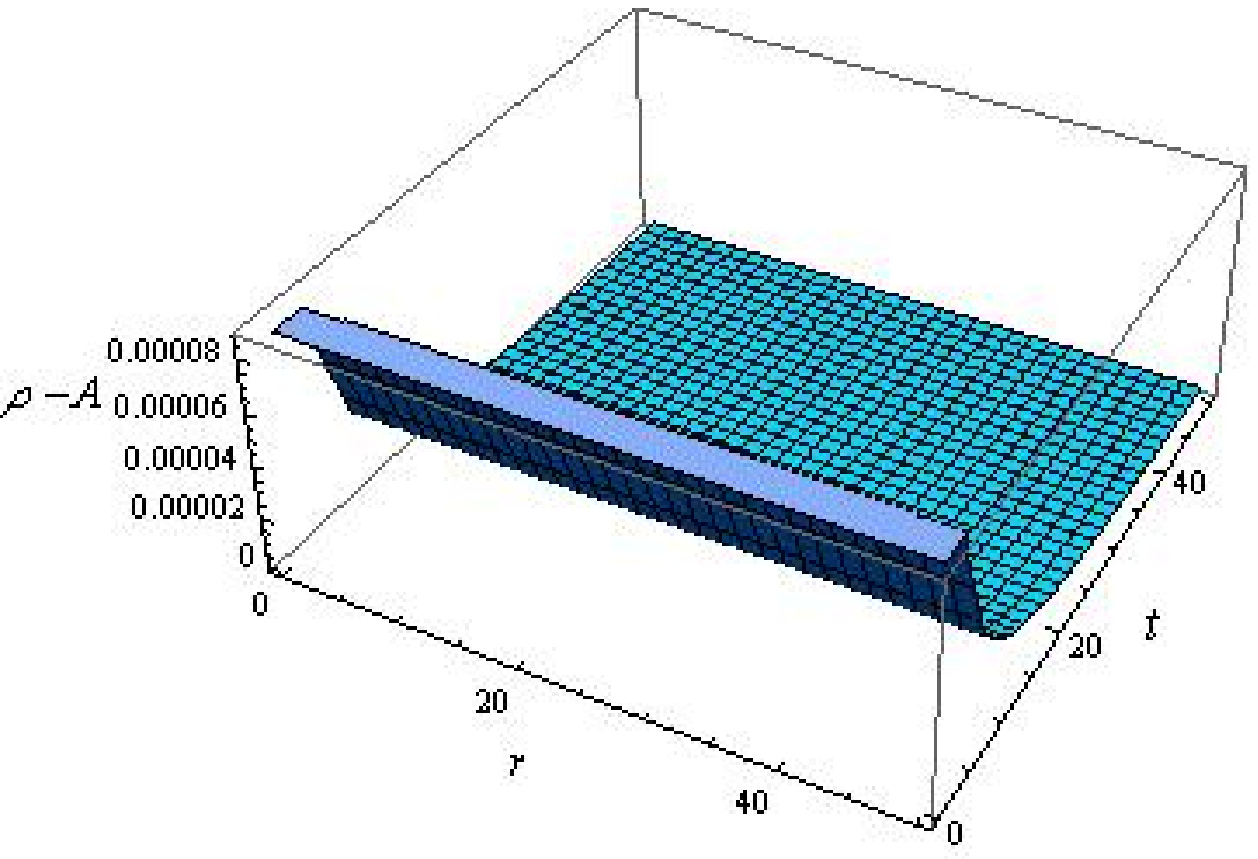,width=0.55\linewidth} \caption{Plot for
WEC for $\gamma=-1.5$, $\alpha=0.1$, $q=0.01$ and $\lambda=-0.1$.}
\end{figure}
\begin{figure}
\center\epsfig{file=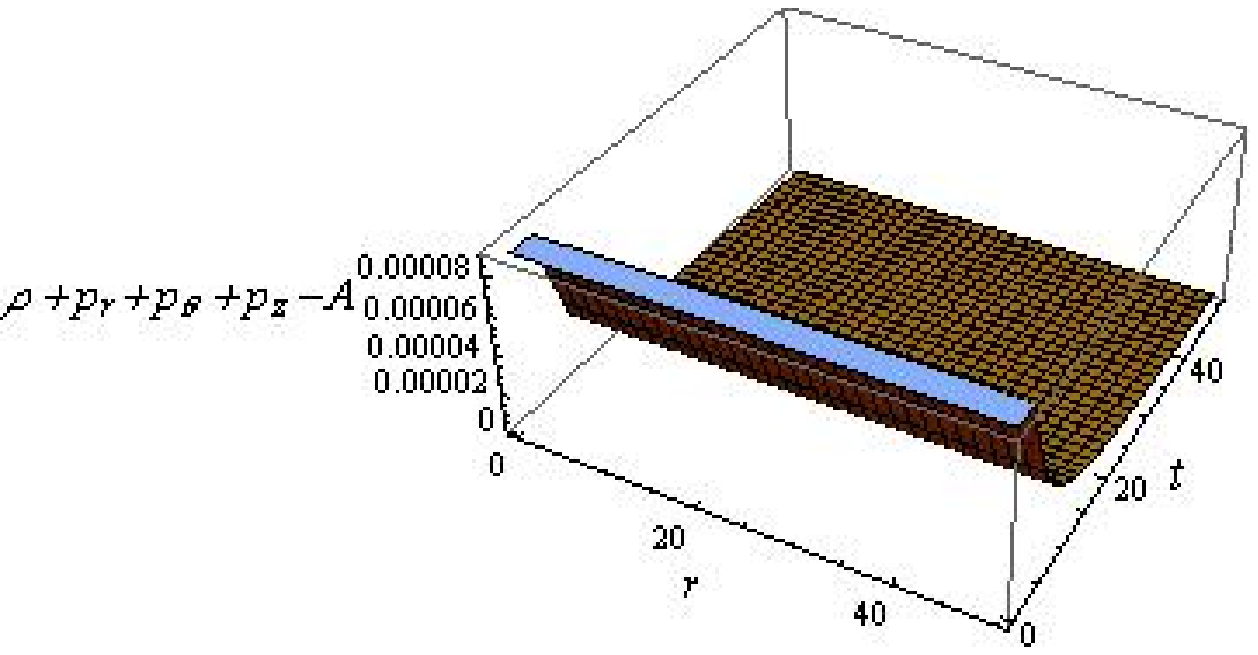,width=0.55\linewidth} \caption{Plot for
SEC for $\gamma=-1.5$, $\alpha=0.1$, $q=0.01$ and $\lambda=-0.1$.}
\end{figure}
\begin{figure}
\epsfig{file=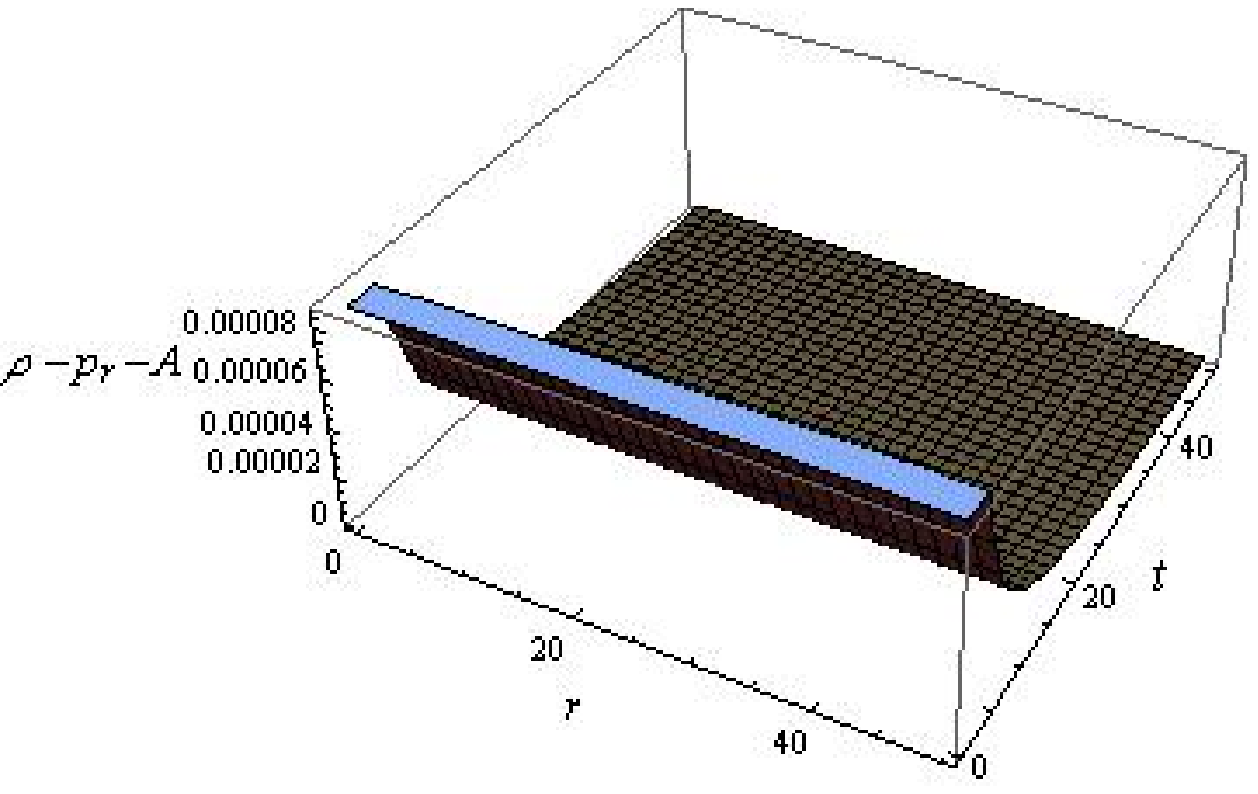,width=0.55\linewidth}\epsfig{file=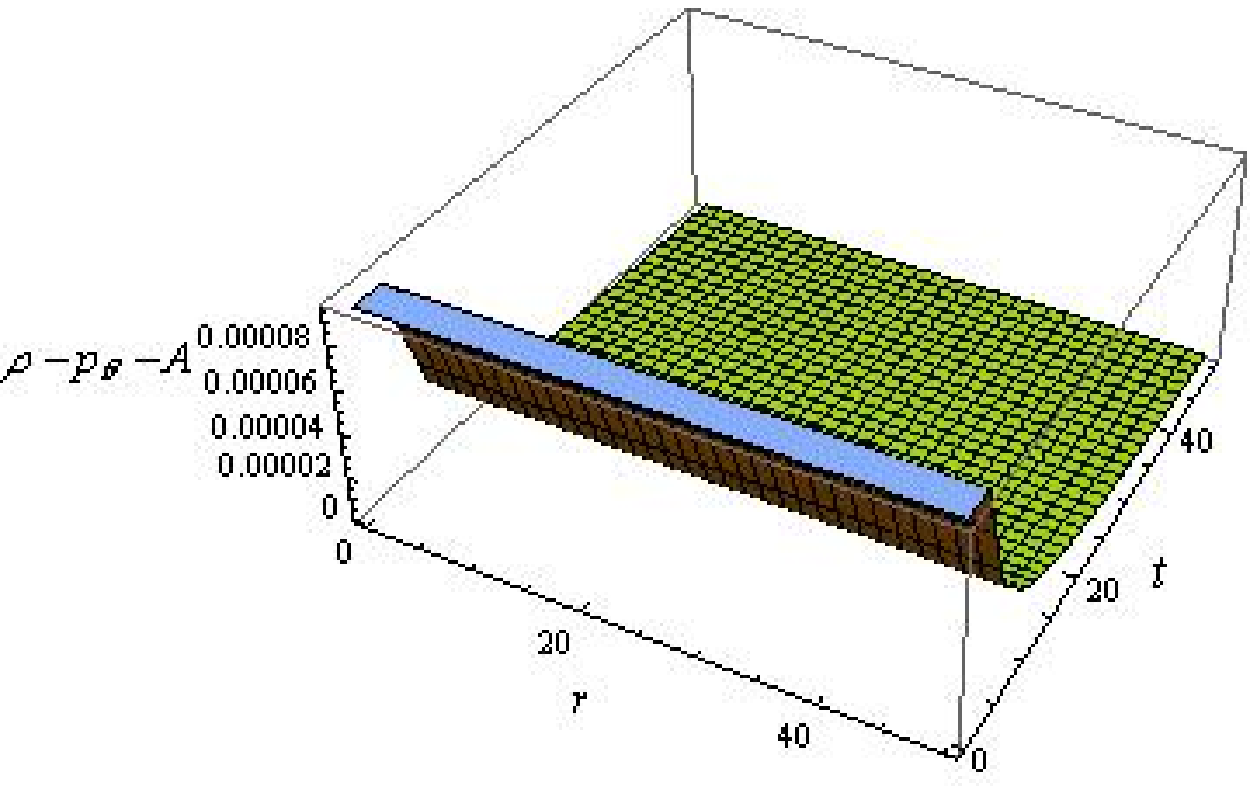,width=0.55\linewidth}
\center\epsfig{file=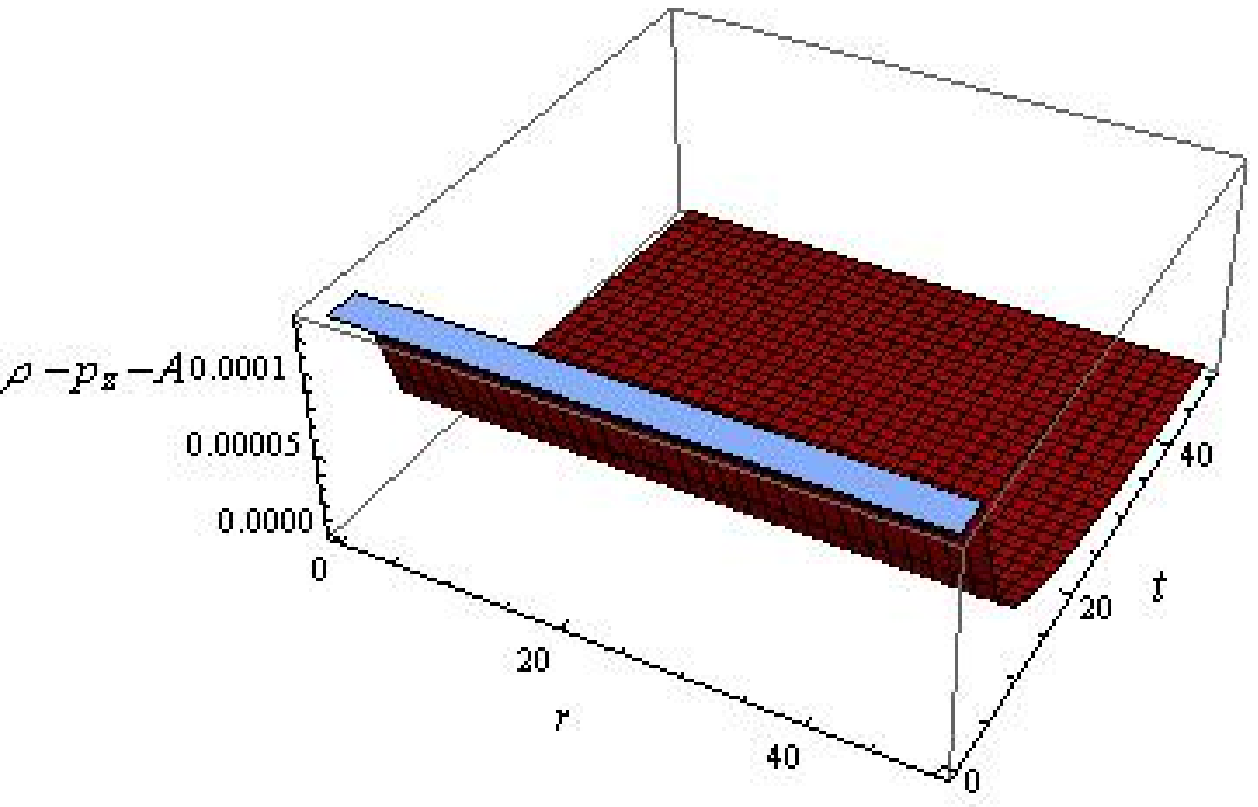,width=0.55\linewidth} \caption{Plots
for DEC for $\gamma=-1.5$, $\alpha=0.1$, $q=0.01$ and
$\lambda=-0.1$.}
\end{figure}

\subsection{Expansion for $\gamma>-1$}
\begin{figure}
\center\epsfig{file=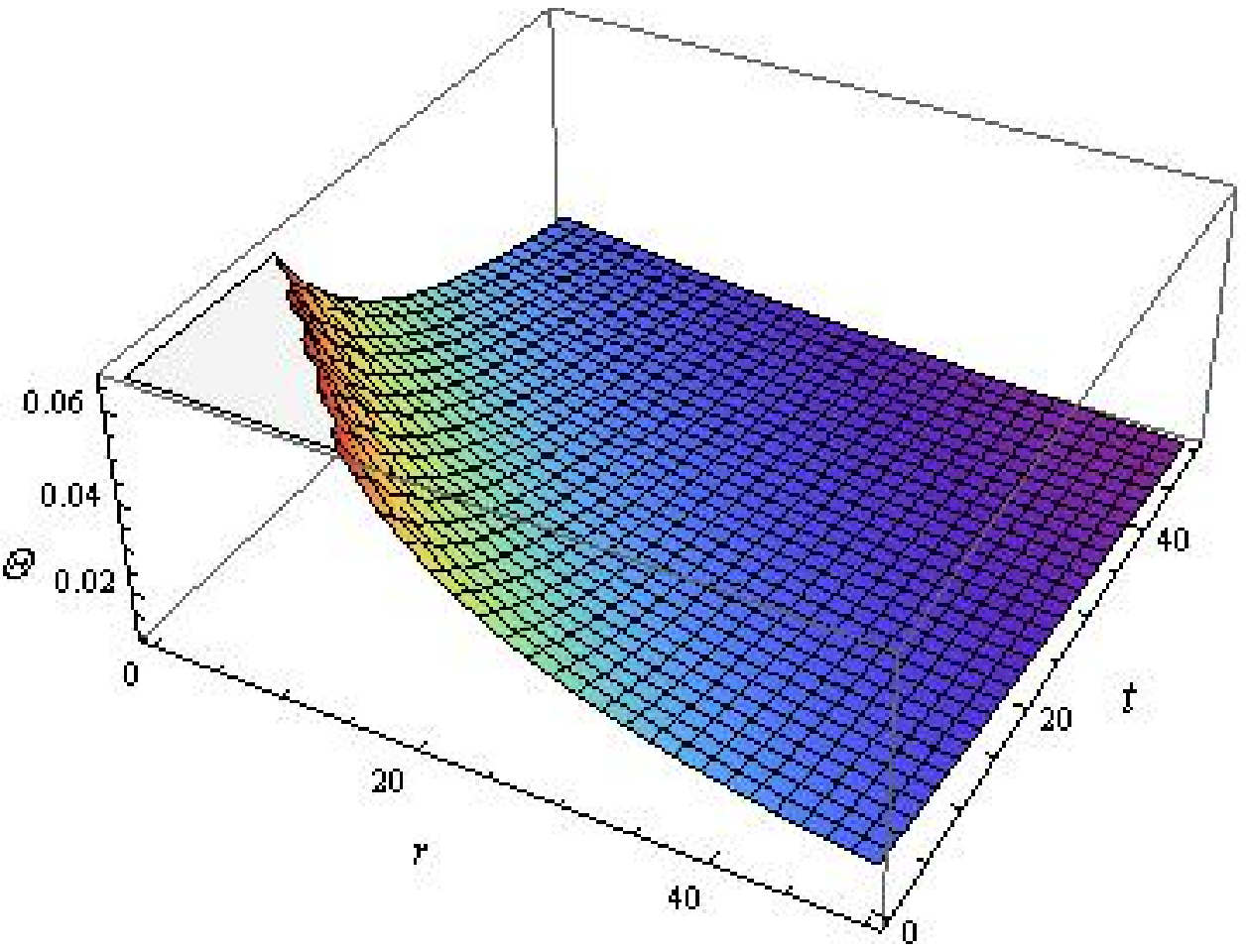,width=0.55\linewidth}\caption{Plot of
$\Theta$ versus $r$ and $t$ for $\gamma=0.0001$, $\alpha=1$.}
\end{figure}

In this case, we require an expression of the metric coefficient $C$
for expanding solution. For convenience, We assume it a linear
combination of $r$ and $t$ such that the expansion scalar remains
positive for the resulting solution as shown in Figure \textbf{12}.
Thus the expanding solution is given by
\begin{eqnarray}\label{30}
A=\frac{1}{\alpha(r+t)^{\gamma}}, \quad B=\alpha(r+t)^{\gamma},\quad
C=r+t.
\end{eqnarray}
Consequently, the expressions of $\rho$, $p_{r}$, $p_{\theta}$ and
$p_{z}$ take the form
\begin{eqnarray}\nonumber
\rho&=&\frac{(r+t)^{-2(2+\gamma)}}{8(8\pi^{2}+6\pi\lambda+\lambda^{2})\alpha^{2}}
\left[8\pi(r+t)^{2}(q^{2}+\gamma+(r+t)^{4\gamma}\gamma
C^{4})\right.\\\nonumber
&+&\left.\lambda(q^{2}(r+t)^{2}(5-2(r+t)^{2\gamma}\alpha^{2}+
(r+t)^{4\gamma}\alpha^{4})-\gamma(-1-2\gamma\right.\\\nonumber
&+&\left.t^{2}(-5-2\gamma-3(r+t)^{4\gamma}\alpha^{4}+
4(r+t)^{4\gamma}\gamma\alpha^{4})+r^{2}(-5\right.\\\label{31}
&-&\left.3(r+t)^{4\gamma}\alpha^{4}
+\gamma(-2+4(r+t)^{4\gamma}\alpha^{4}))))\right],\\\nonumber
p_{r}&=&\frac{1}{8(r+t)^{2}(2\pi+\lambda)(4\pi+\lambda)}
\left[\frac{-q^{2}(r+t)^{-2\gamma}\lambda}{\alpha^{2}}
-\frac{(r+t)^{-2(1+\gamma)}\gamma}{\alpha^{2}}\right.\\\nonumber
&\times&\left.
(8\pi(r+t)^{2}+(1+2\gamma+r^{2}(5+2\gamma)+2rt(5+2\gamma)
+t^{2}(5+2\gamma))\lambda)\right.\\\nonumber&+&\left.2(r+t)^{2\gamma}(-1+\gamma)\gamma\lambda\alpha^{2}+
(r+t)^{2\gamma}\gamma(-16\pi+(-5+2\gamma)\lambda)\alpha^{2}\right.\\\label{32}&+&\left.
q^{2}(2\lambda+(r+t)^{2\gamma}(8\pi+3\lambda)\alpha^{2})\right],\\\nonumber
p_{\theta}&=&\frac{1}{8(r+t)^{2}(2\pi+\lambda)(4\pi+\lambda)}
\left[\frac{-q^{2}(r+t)^{-2\gamma}\lambda}{\alpha^{2}}+
\frac{(r+t)^{-2(1+\gamma)}\gamma(1+2\gamma)}{\alpha^{2}}\right.\\\nonumber&\times&\left.
(-8\pi+(-3+r^{2}+2rt+t^{2})\lambda)+2(r+t)^{2\gamma}(-1+\gamma)\gamma(4\pi+\lambda)\alpha^{2}
\right.\\\label{33}&+&\left.(r+t)^{2\gamma}\gamma(8\pi\gamma+(-3+2\gamma)\lambda)\alpha^{2}
+q^{2}(8\pi+\lambda(2+(r+t)^{2\gamma}\alpha^{2}))\right],\\\nonumber
p_{z}&=&\frac{(r+t)^{-2(2+\gamma)}}{8(2\pi+\lambda)(4\pi+\lambda)\alpha^{2}}
\left[-8\pi(r+t)^{2}(q^{2}(r+t)^{2\gamma}\alpha^{2}+
(\gamma+2\gamma^{2})(-1\right.\\\nonumber&+&\left.(r+t)^{4\gamma}\alpha^{4})-
\lambda(q^{2}(r+t)^{2}(1+(r+t)^{2\gamma}\alpha^{2})^{2}+\gamma(1+2\gamma+t^{2}\right.\\\nonumber&\times&\left.
(-3-6\gamma+3(r+t)^{4\gamma}\alpha^{4}+4(r+t)^{4\gamma}\gamma\alpha^{4})
+(r^{2}+2rt)(-3\right.\\\label{34}&+&\left.
3(r+t)^{4\gamma}\alpha^{4}+\gamma(-6+4(r+t)^{4\gamma}\alpha^{4})))))\right].
\end{eqnarray}
The anisotropic parameter and mass function are obtained as
\begin{eqnarray}\nonumber
\triangle&=&\frac{(r+t)^{-2(2+\gamma)}}{2(4\pi+\lambda)\alpha^{2}}
\left[q^{2}(r+t)^{2+2\gamma}\alpha^{2}(1+(r+t)^{2\gamma}\alpha^{2})
+2\gamma^{2}\right.\\\nonumber&\times&\left.(-1+(r^{2}+2rt+t^{2})(r+t)^{4\gamma}\alpha^{4})
-\gamma(1+t^{2}+3t^{2}(r+t)^{4\gamma}\alpha^{4}
\right.\\\label{36}&+&\left.(r^{2}+2rt)(1+3(r+t)^{4\gamma}\alpha^{4}))\right],\\\label{37}
m&=&\frac{1}{8}+q(r+t).
\end{eqnarray}
The evolution of physical parameters during expansion is represented
through Figures \textbf{13}-\textbf{18}. It is found that the
quantities vary with both time and radial coordinates. The graphical
analysis is summarized in Tables \textbf{3} and \textbf{4}.\\\\
Table \textbf{3}: Change in parameters with respect to $r$ and $t$
for the expanding solution.
\begin{table}[bht]
\centering
\begin{small}
\begin{tabular}{|c|c|c|c|c|c|c|}
\hline \textbf{Parameter} &$\rho$&$p_{r}$&$p_{\theta}$&$p_{z}$&$\triangle$&$m$\\
\hline \textbf{As $r$ increases}&decreases&decreases&increases&increases&decreases&increases\\
\hline \textbf{As $t$ increases}&decreases&decreases&increases&increases&decreases&increases\\
\hline
\end{tabular}
\end{small}
\end{table}\\
Table \textbf{4}: Effects of $q$ and $\lambda$ for the expanding
solution.
\begin{table}[bht]
\centering
\begin{small}
\begin{tabular}{|c|c|c|c|c|c|c|}
\hline\textbf{Parameter}&$\rho$&$p_{r}$&$p_{\theta}$&$p_{z}$&$\triangle$&$m$\\
\hline\textbf{As $q$ increases}&increases&increases&decreases&decreases&increases&increases\\
\hline\textbf{As $\lambda$ decreases}&increases&increases&decreases&decreases&increases&no change\\
\hline
\end{tabular}
\end{small}
\end{table}\\
The acceleration term $\mathcal{A}$ in this case becomes
\begin{equation}\nonumber
\mathcal{A}=\frac{(r+t)^{-2(1+\gamma)}\gamma^{2}}{\alpha^{2}}(3+(r+t)^{4\gamma}\alpha^{4}).
\end{equation}
The graphs for energy conditions for expanding solutions are given
in Figures \textbf{19}-\textbf{22} showing that all the energy
conditions are satisfied.
\begin{figure}
\epsfig{file=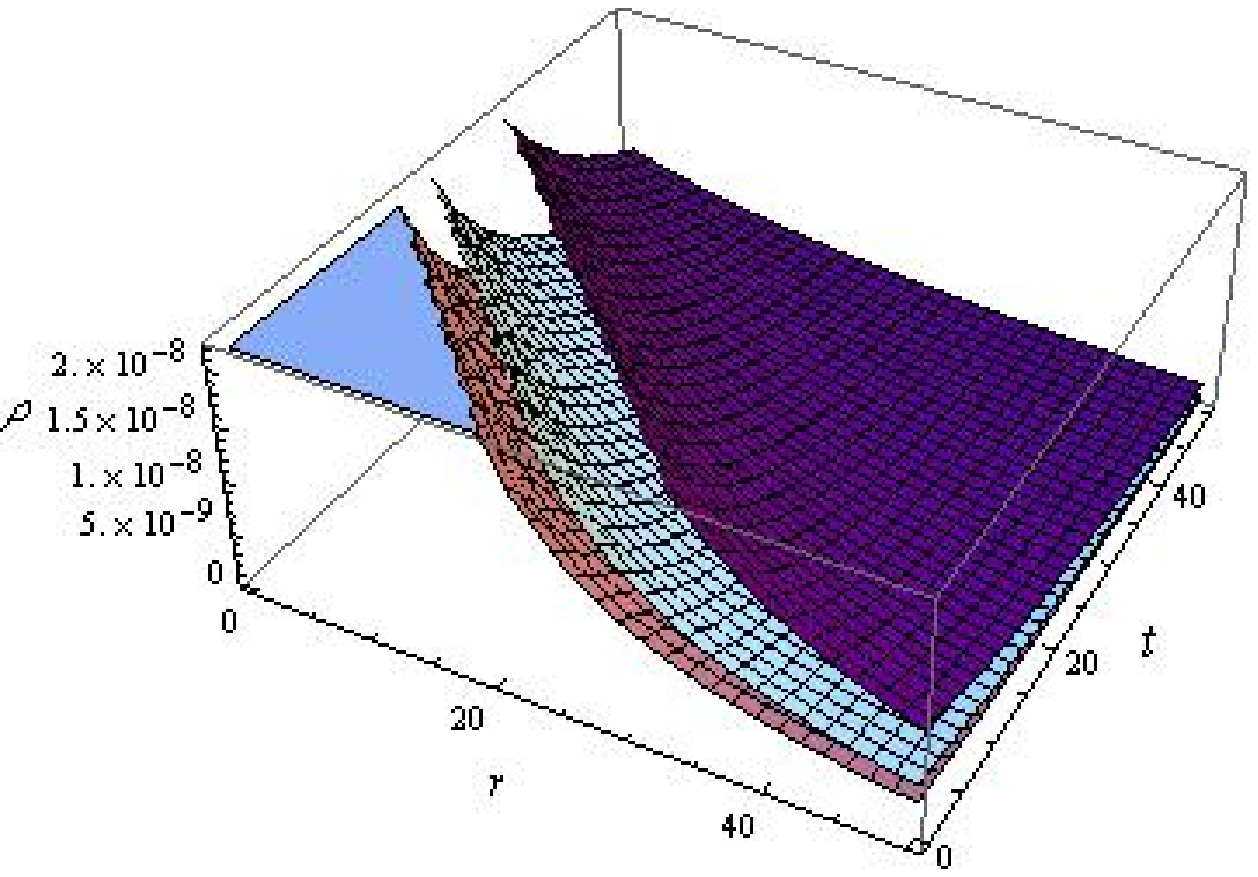,width=0.55\linewidth}\epsfig{file=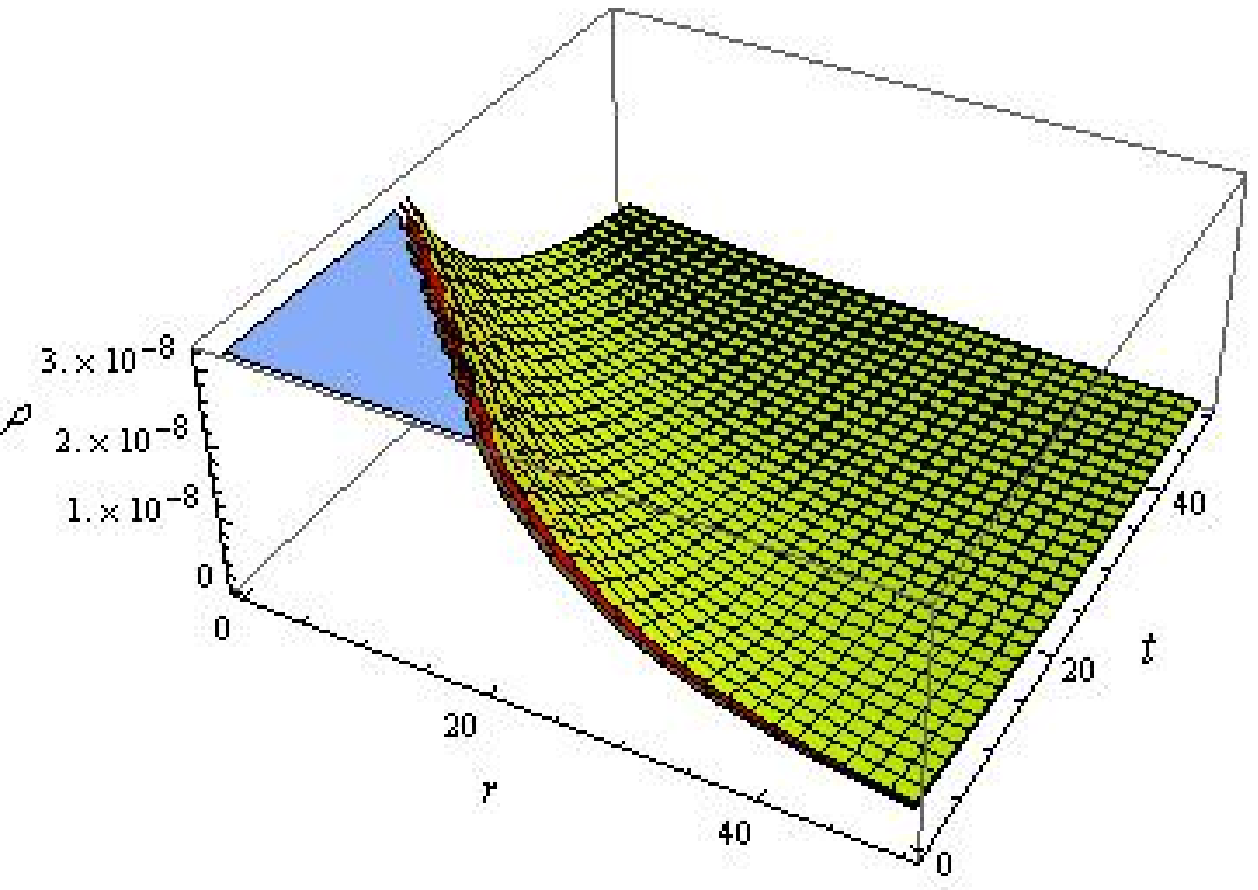,width=0.55\linewidth}
\caption{Plots of $\rho$ versus $r$ and $t$ for $\gamma=0.0001$,
$\alpha=1$. The left graph is for $q=0$ (pink), $q=0.01$ (blue),
$q=0.02$ (purple) with $\lambda=-0.001$ and the right graph is for
$\lambda=-0.001$ (brown), $\lambda=-1$ (red), $\lambda=-2$ (yellow)
with $q=0.01$.}
\end{figure}
\begin{figure}
\epsfig{file=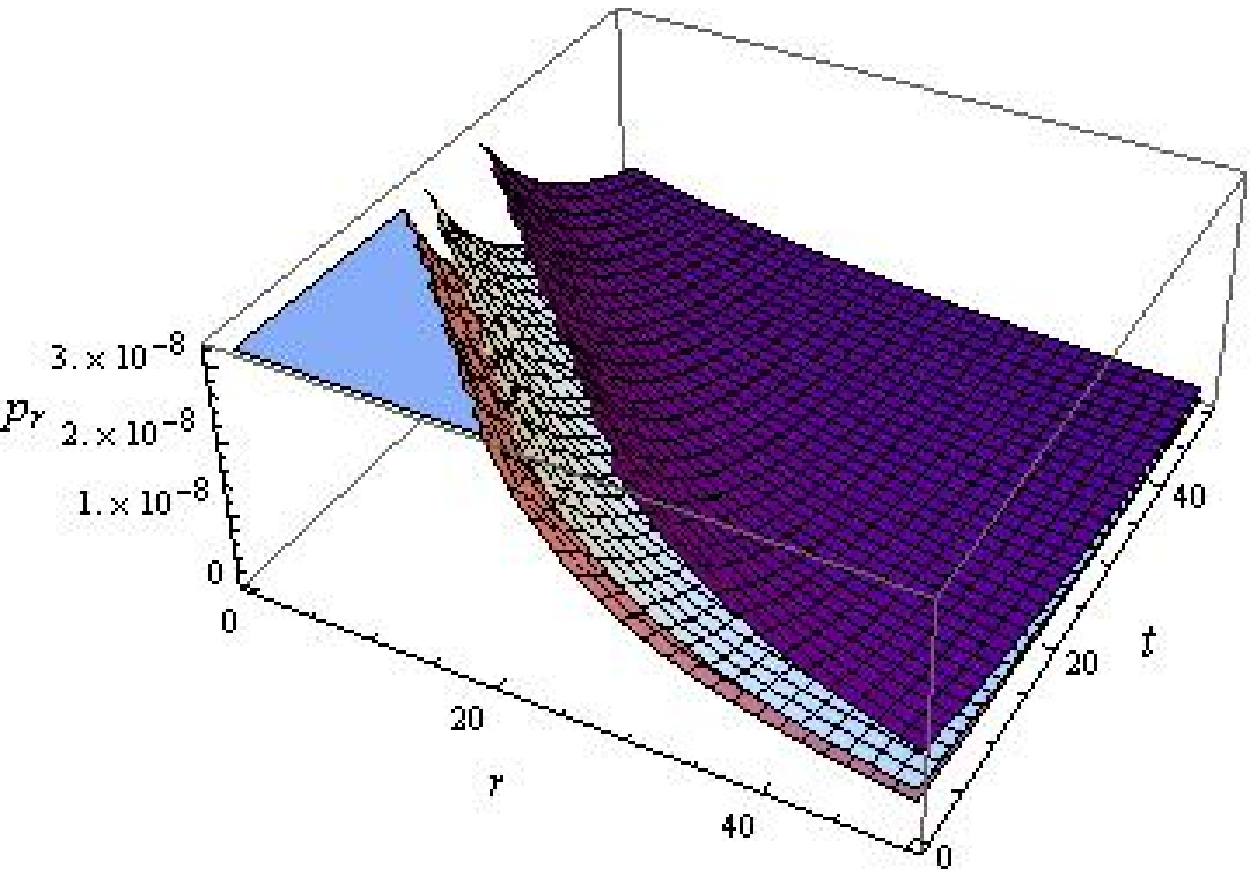,width=0.55\linewidth}\epsfig{file=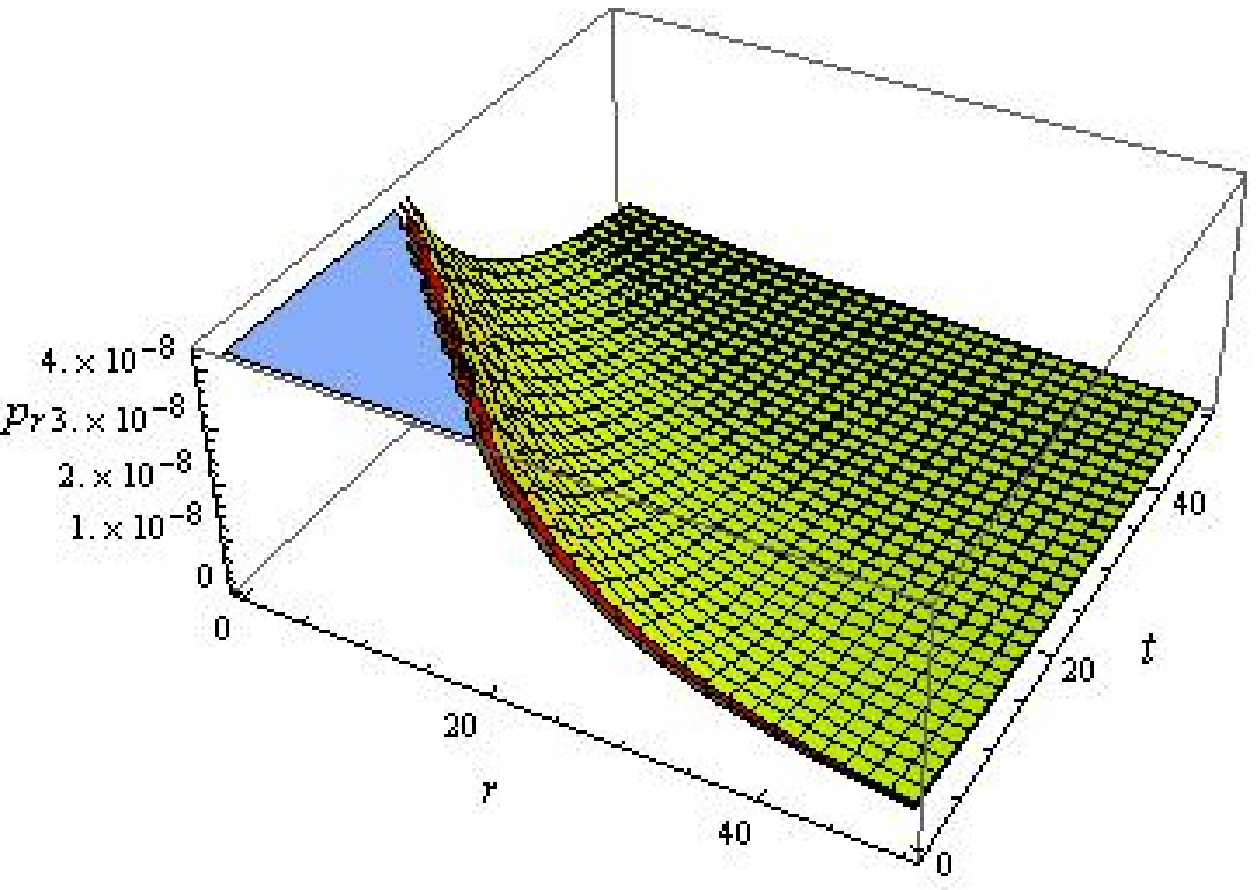,width=0.55\linewidth}
\caption{Plots of $p_{r}$ versus $r$ and $t$ for $\gamma=0.0001$,
$\alpha=1$. The left graph is for $q=0$ (pink), $q=0.01$ (blue),
$q=0.02$ (purple) with $\lambda=-0.001$ while the right graph is for
$\lambda=-0.001$ (brown), $\lambda=-1$ (red), $\lambda=-2$ (yellow)
with $q=0.01$.}
\end{figure}
\begin{figure}
\epsfig{file=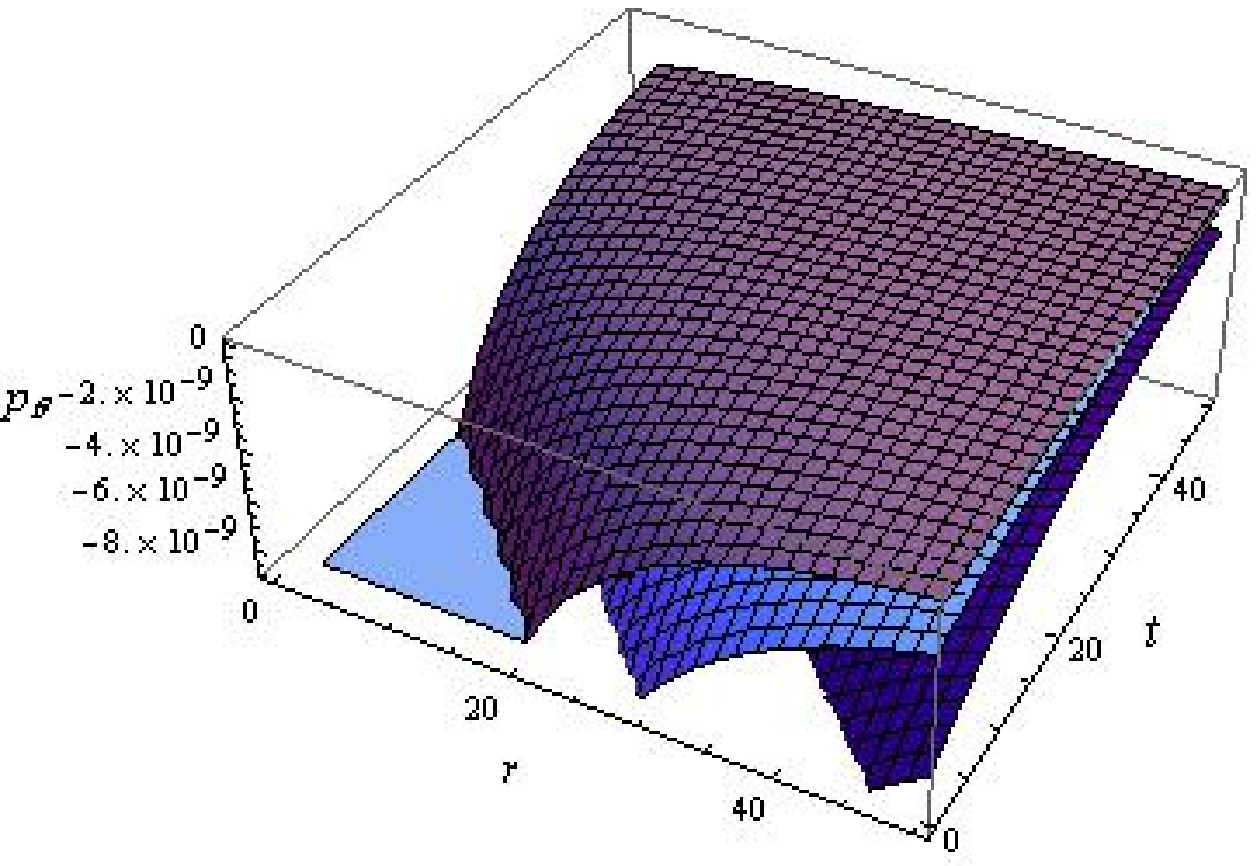,width=0.55\linewidth}\epsfig{file=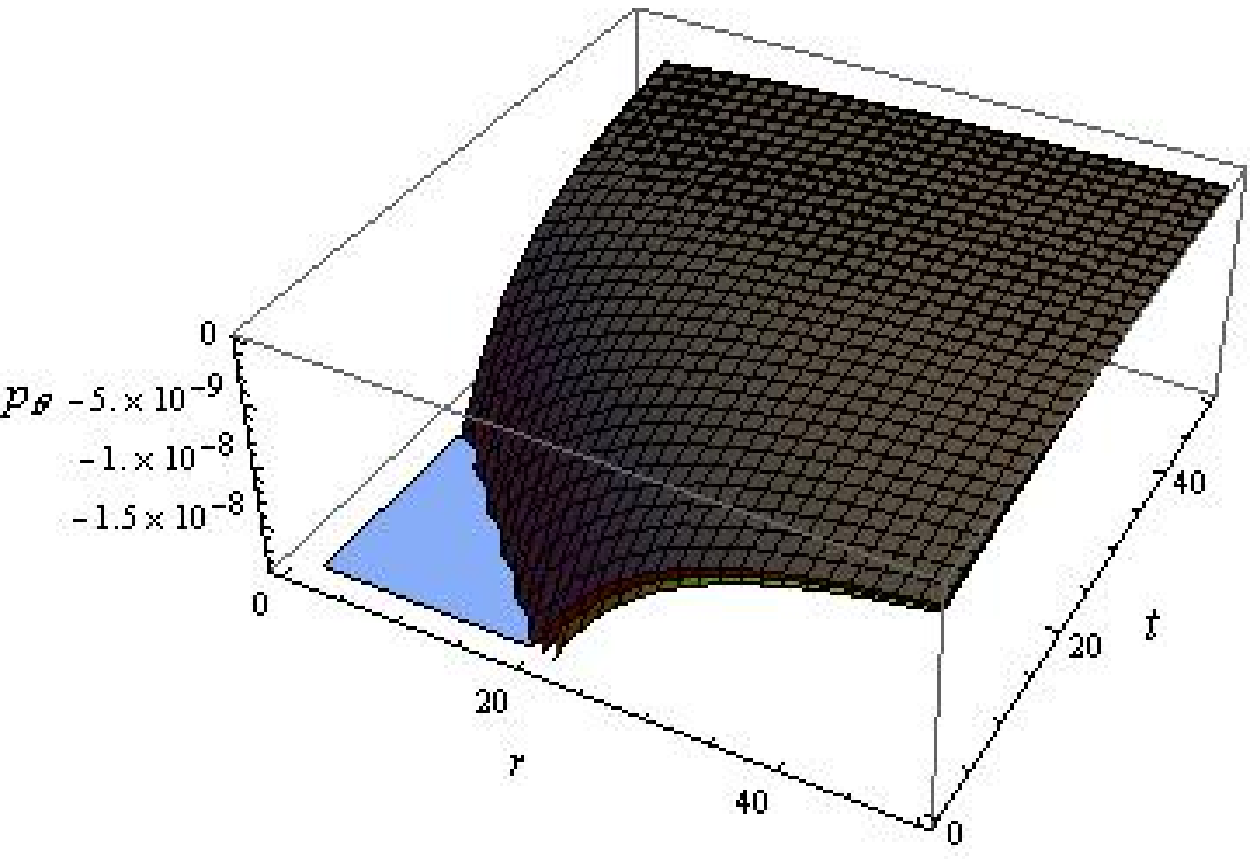,width=0.55\linewidth}
\caption{Plots of $p_{\theta}$ versus $r$ and $t$ for
$\gamma=0.0001$, $\alpha=1$. The left graph is for $q=0$ (pink),
$q=0.01$ (blue), $q=0.02$ (purple) with $\lambda=-0.001$ while the
right graph is for $\lambda=-0.001$ (brown), $\lambda=-1$ (red),
$\lambda=-2$ (yellow) with $q=0.01$.}
\end{figure}
\begin{figure}
\epsfig{file=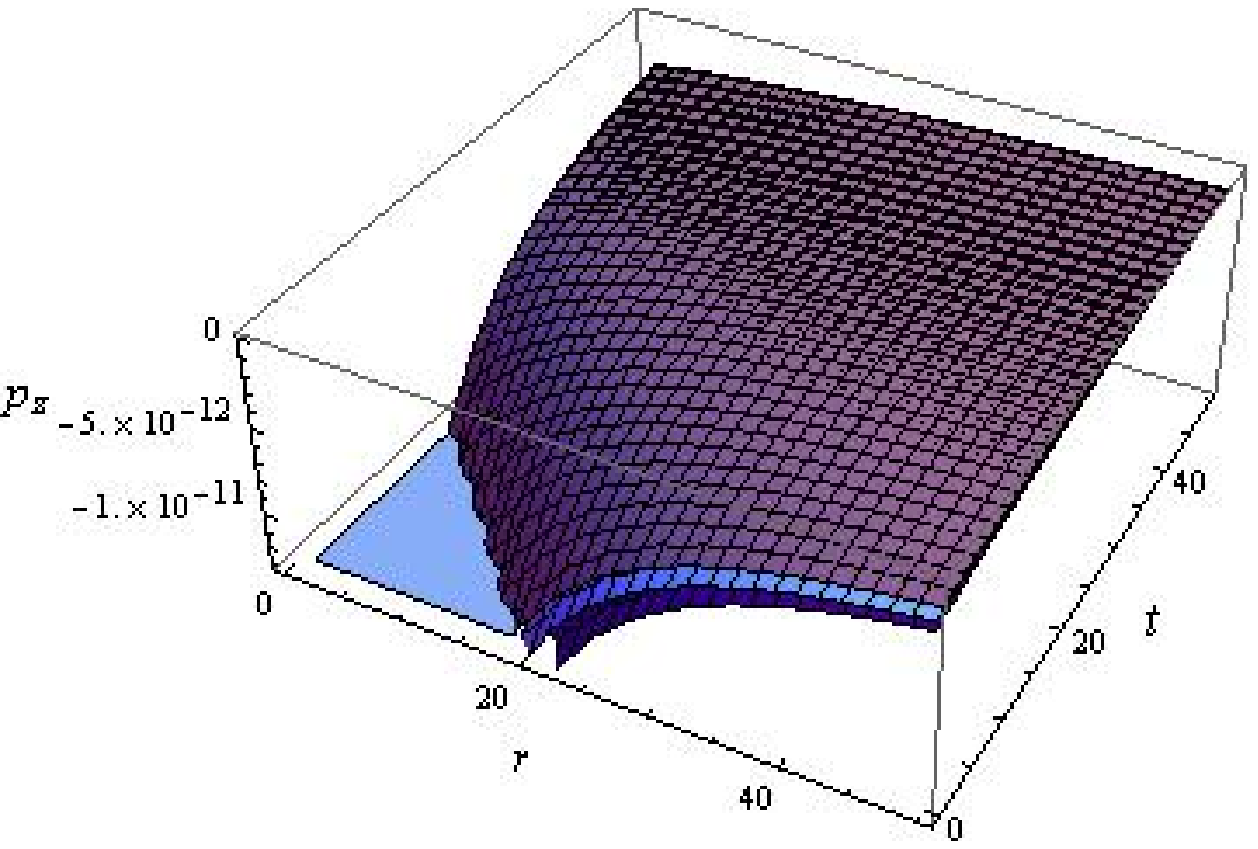,width=0.55\linewidth}\epsfig{file=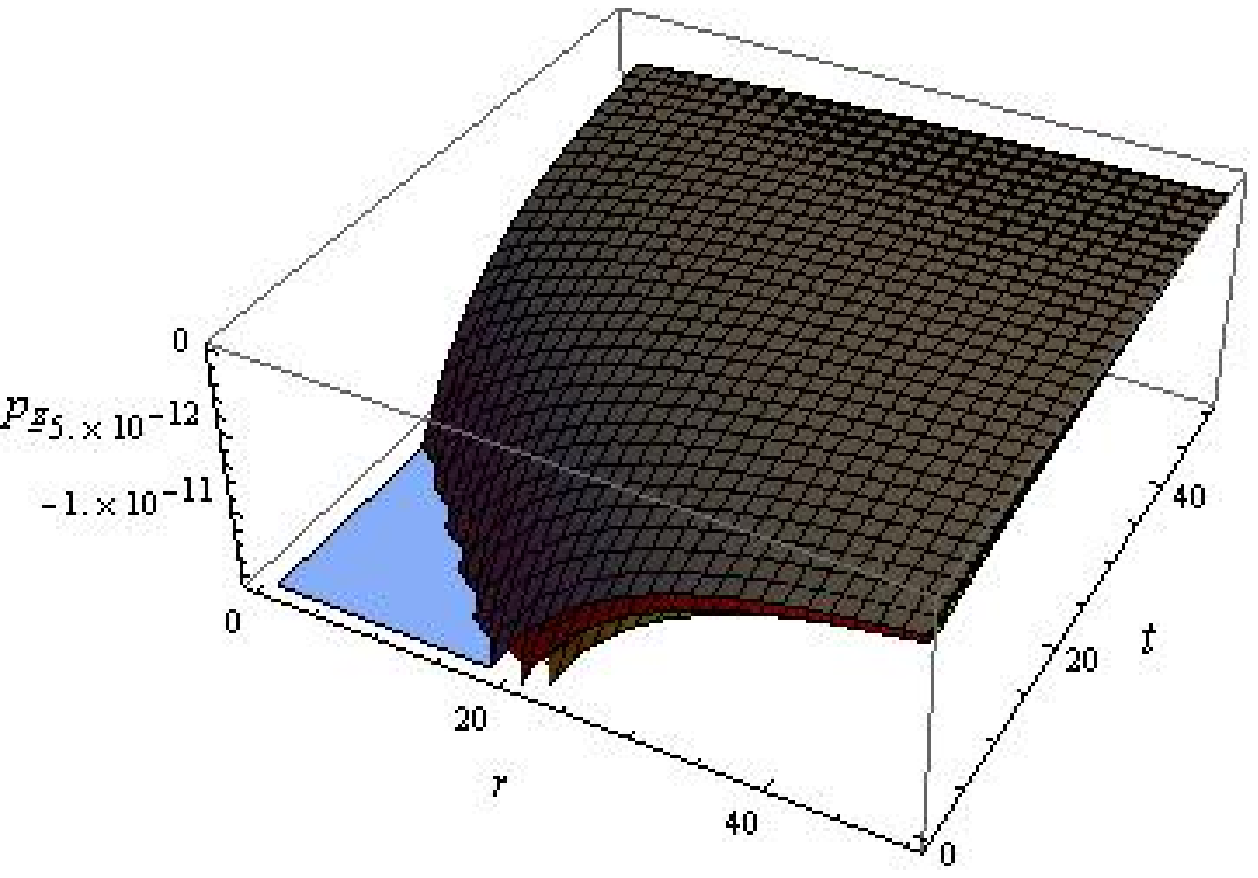,width=0.55\linewidth}
\caption{Plots of $p_{z}$ versus $r$ and $t$ for $\gamma=0.0001$,
$\alpha=1$. The left graph is for $q=0$ (pink), $q=0.0001$ (blue),
$q=0.0002$ (purple) with $\lambda=-0.001$ while the right graph is
for $\lambda=-0.001$ (brown), $\lambda=-1$ (red), $\lambda=-2$
(yellow) with $q=0.01$.}
\end{figure}
\begin{figure}
\epsfig{file=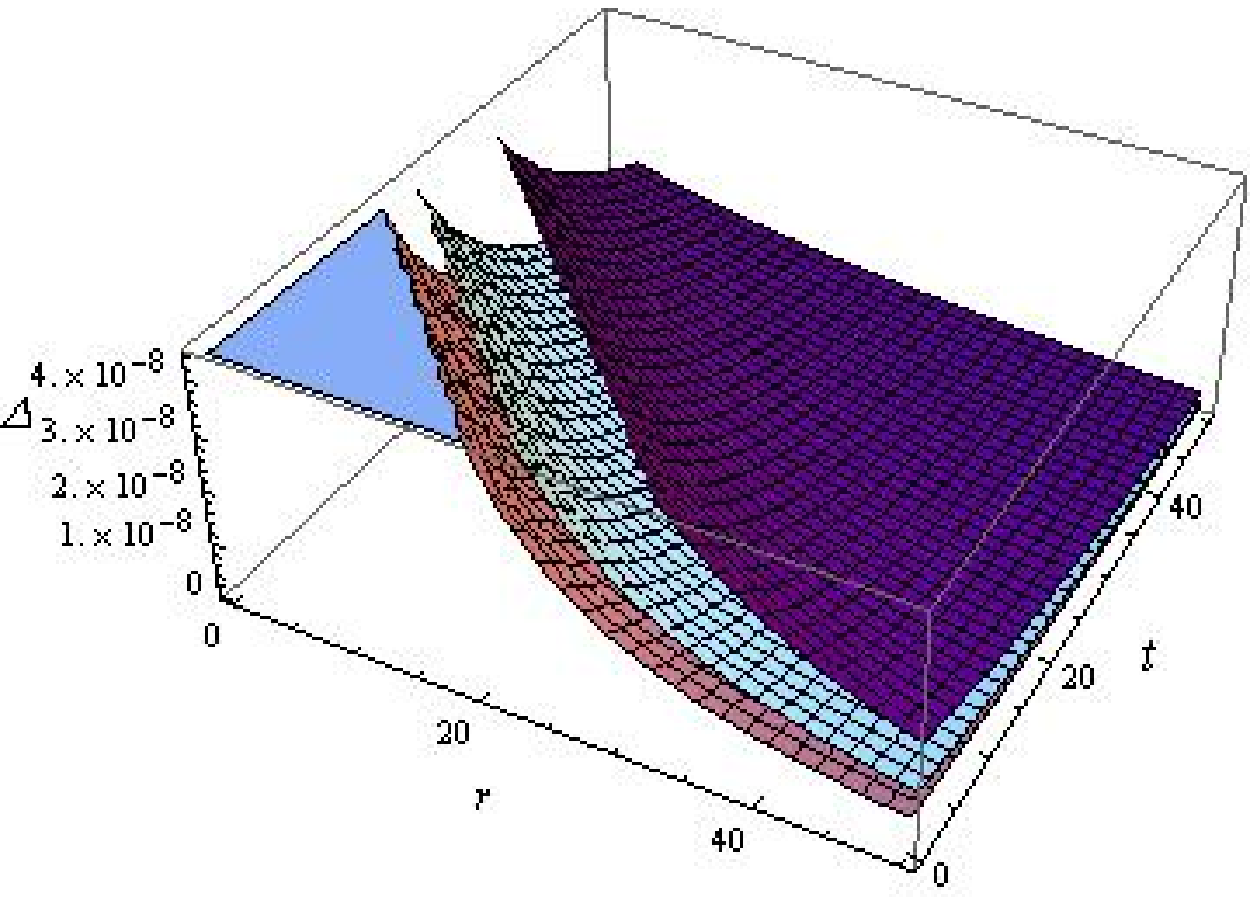,width=0.55\linewidth}\epsfig{file=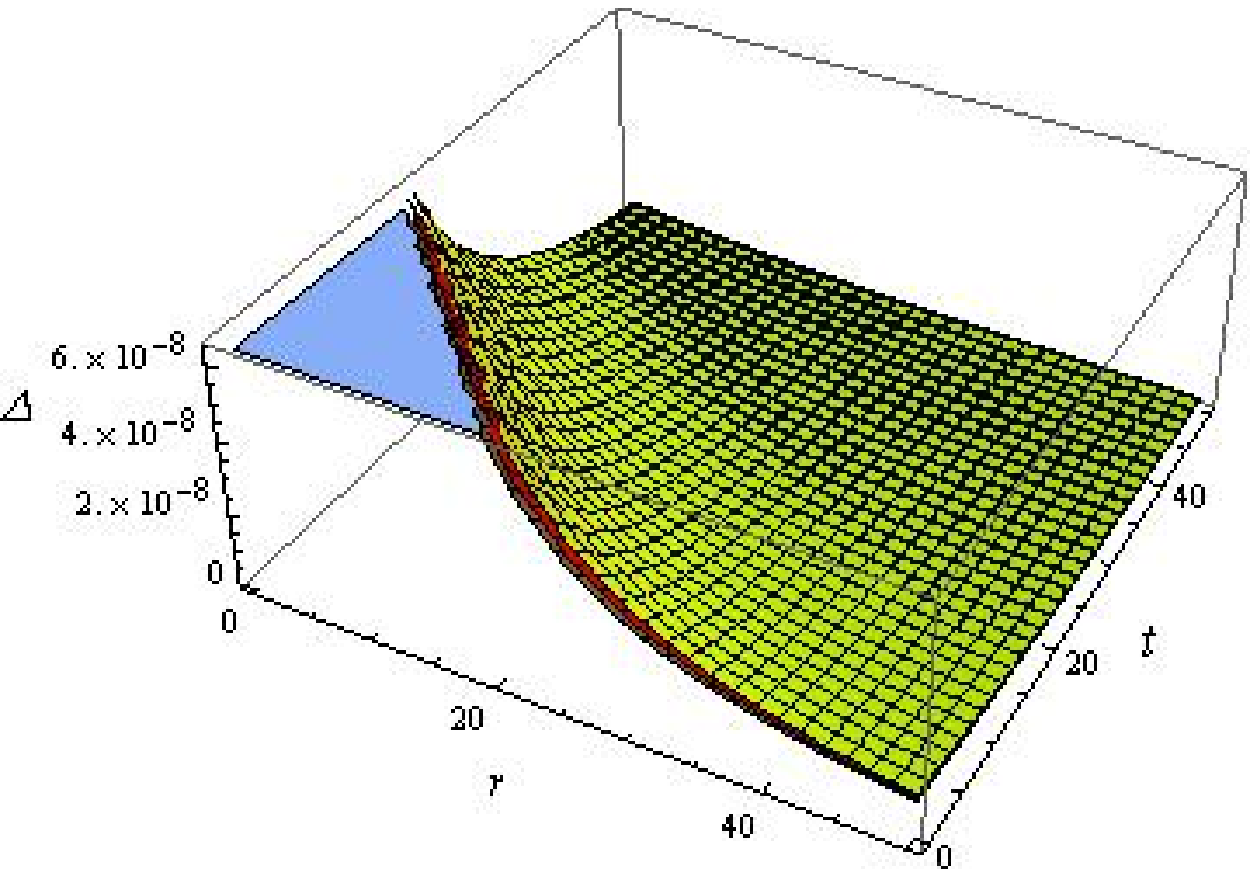,width=0.55\linewidth}\caption{Plots
of $\triangle$ versus $r$ and $t$ for $\gamma=0.0001$, $\alpha=1$.
The left graph is for $q=0$ (pink), $q=0.01$ (blue), $q=0.02$
(purple) with $\lambda=-0.001$ while the right graph is for
$\lambda=-0.001$ (brown), $\lambda=-1$ (red), $\lambda=-2$ (yellow)
with $q=0.5$.}
\end{figure}
\begin{figure}
\center\epsfig{file=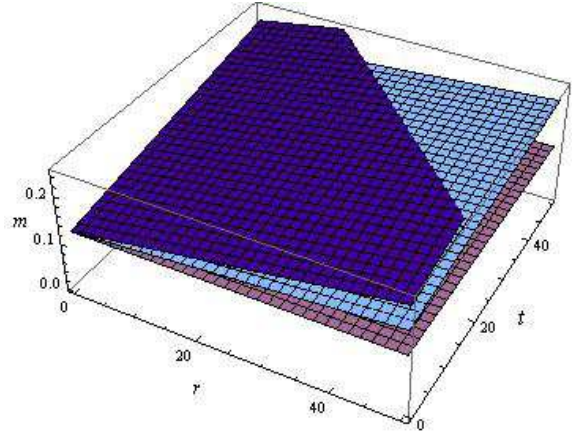,width=0.55\linewidth}\caption{Plot of $m$
versus $r$ and $t$ for $\gamma=0.0001$, $\alpha=1$, $q=0$ (pink),
$q=0.001$ (blue), $q=0.002$ (purple).}
\end{figure}
\begin{figure}
\epsfig{file=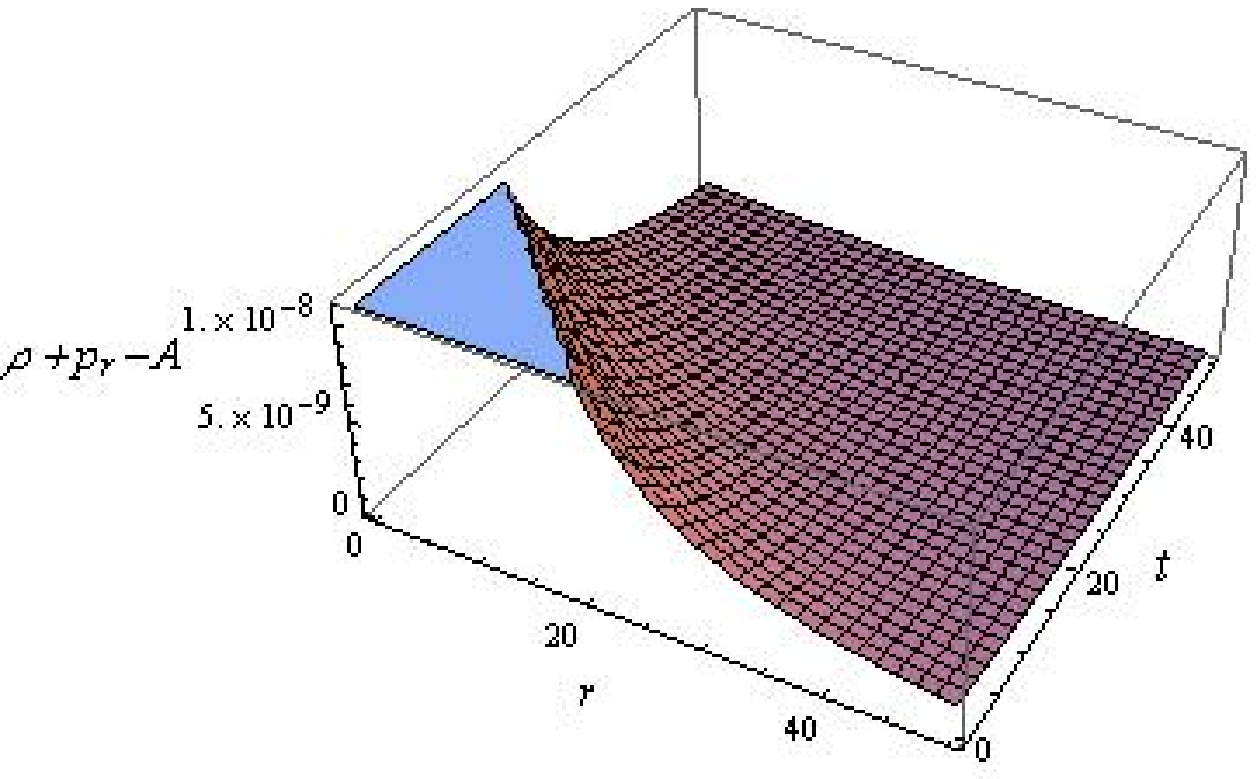,width=0.55\linewidth}\epsfig{file=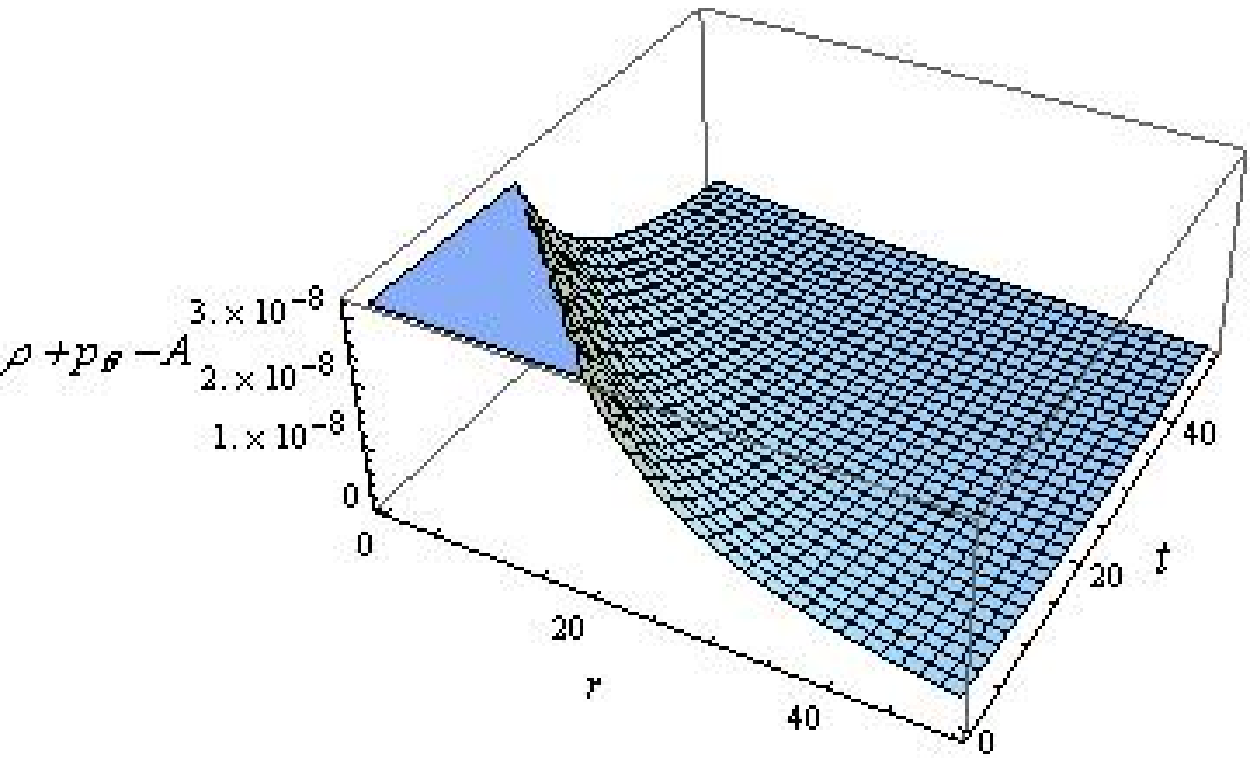,width=0.55\linewidth}
\center\epsfig{file=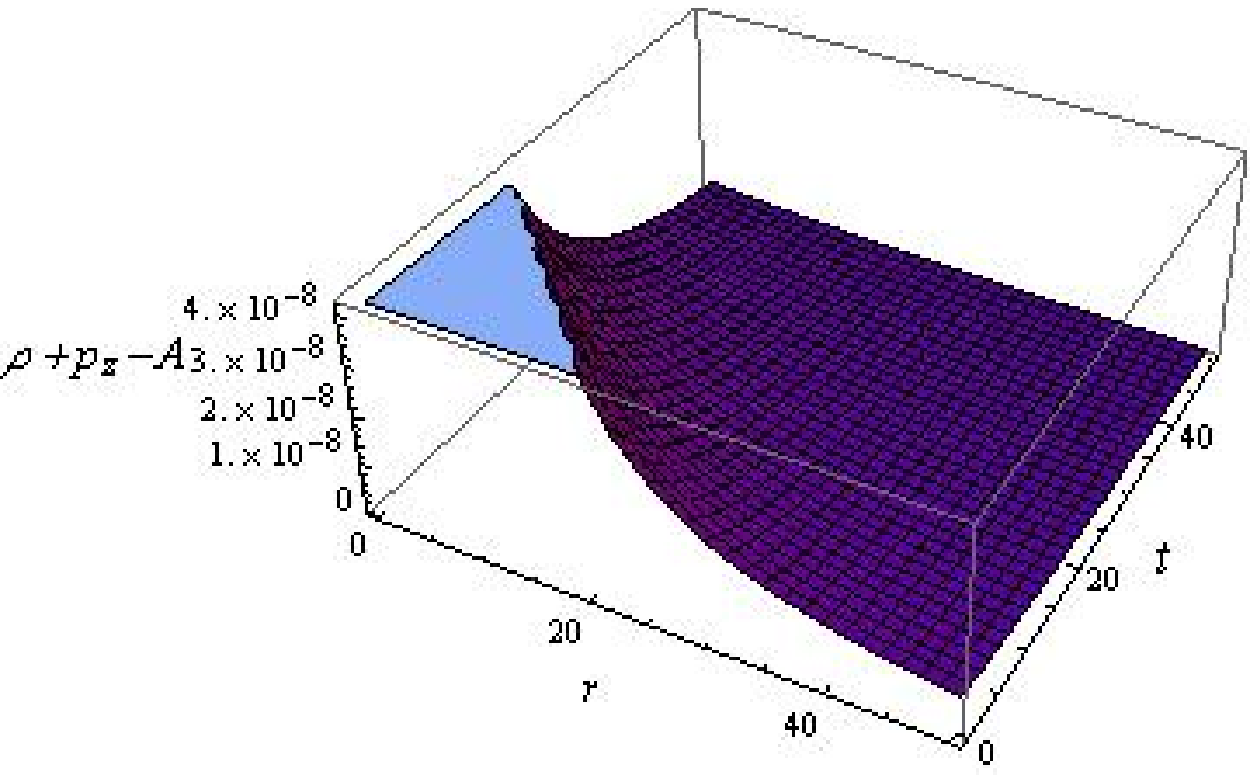,width=0.55\linewidth} \caption{Plots
for NEC for $\gamma=0.0001$, $\alpha=1$, $q=0.01$ and
$\lambda=-0.001$.}
\end{figure}
\begin{figure}
\center\epsfig{file=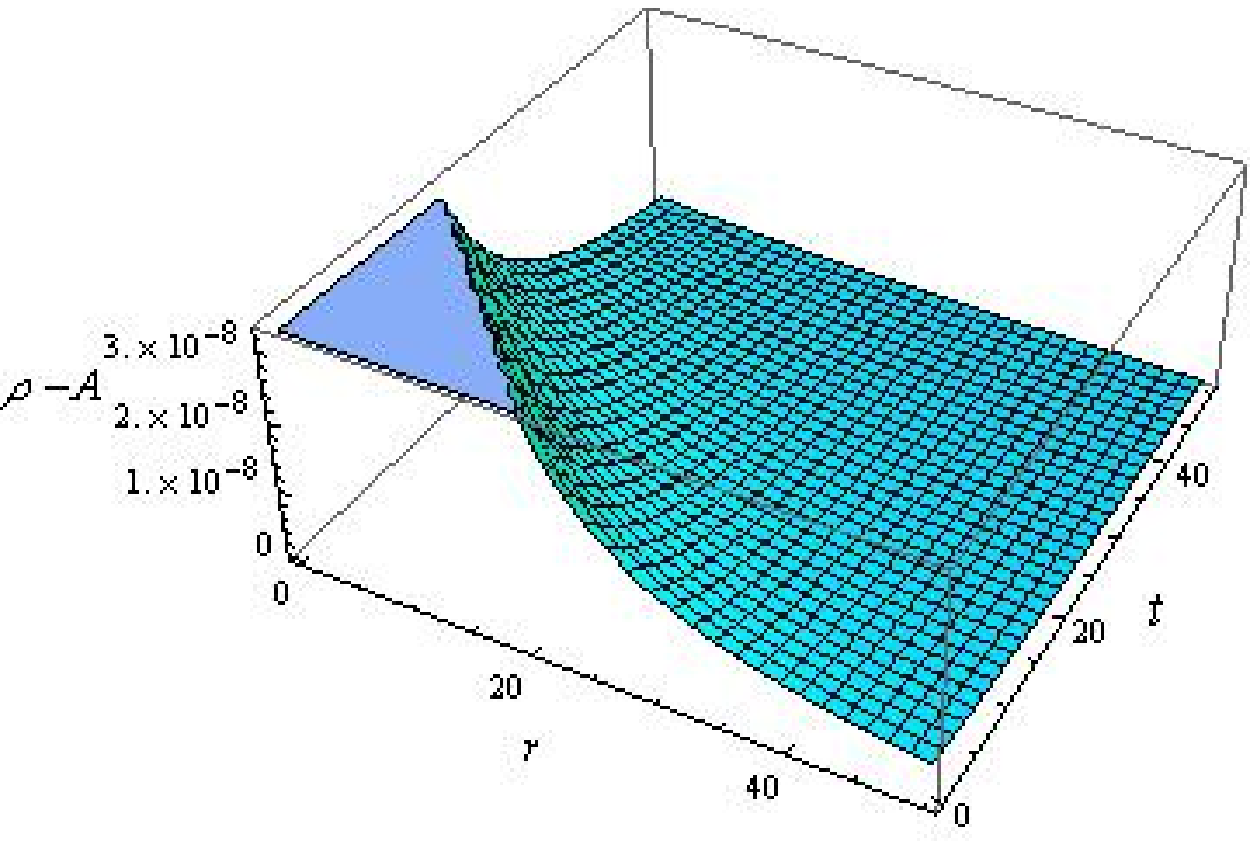,width=0.55\linewidth} \caption{Plot for
WEC for $\gamma=0.0001$, $\alpha=1$, $q=0.01$ and $\lambda=-0.001$.}
\end{figure}
\begin{figure}
\center\epsfig{file=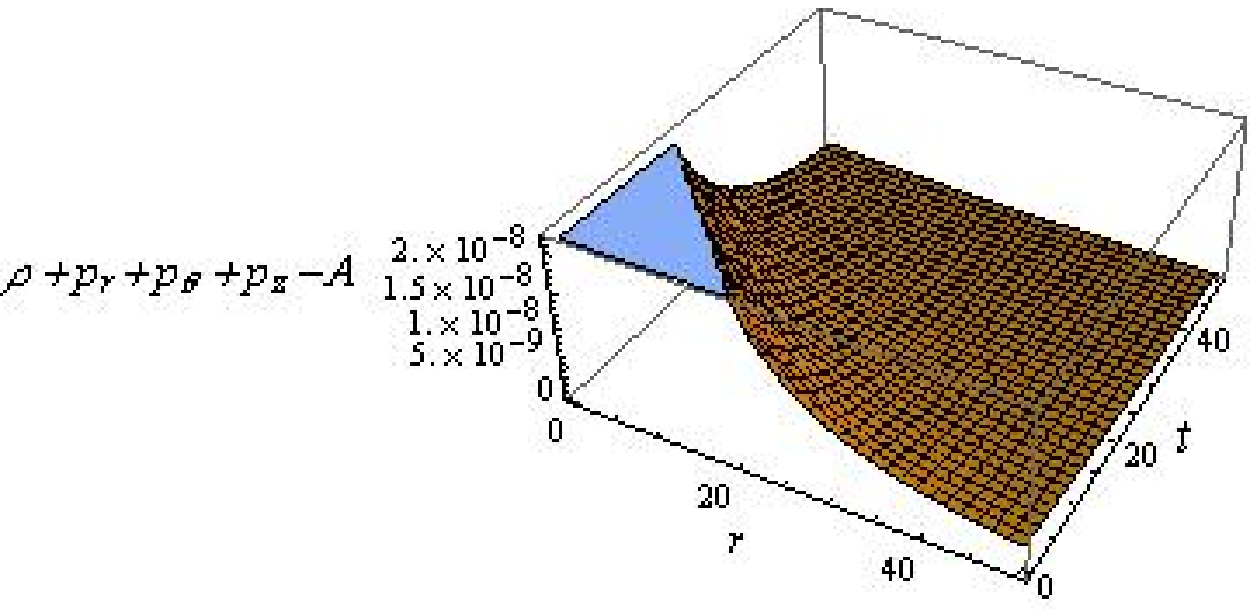,width=0.55\linewidth} \caption{Plot for
SEC for $\gamma=0.0001$, $\alpha=1$, $q=0.01$ and $\lambda=-0.001$.}
\end{figure}
\begin{figure}
\epsfig{file=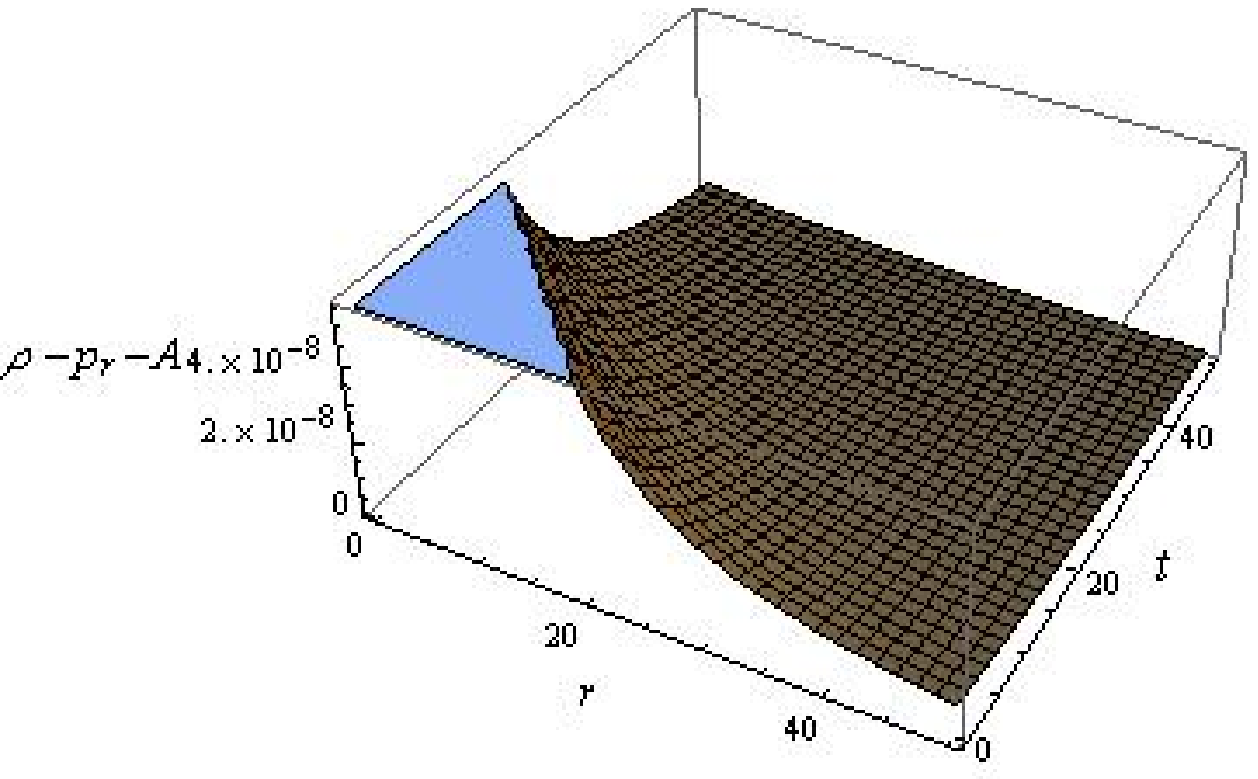,width=0.55\linewidth}\epsfig{file=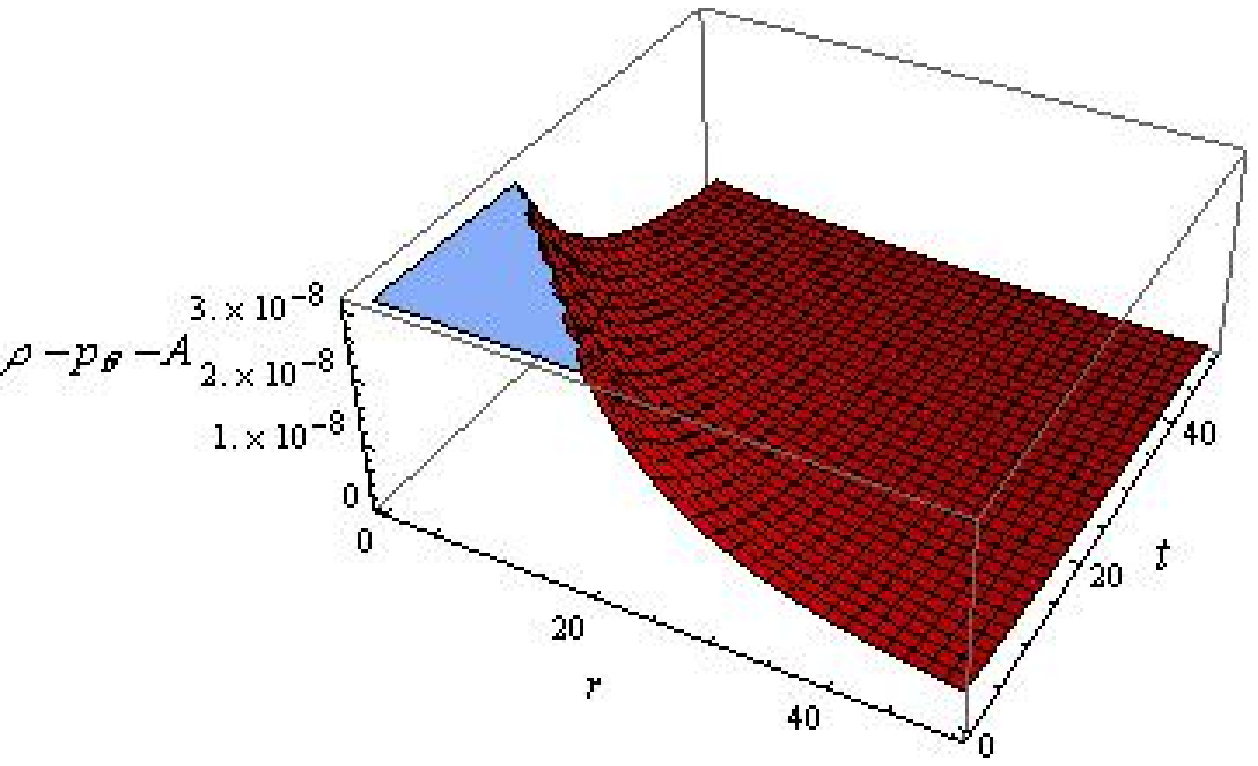,width=0.55\linewidth}
\center\epsfig{file=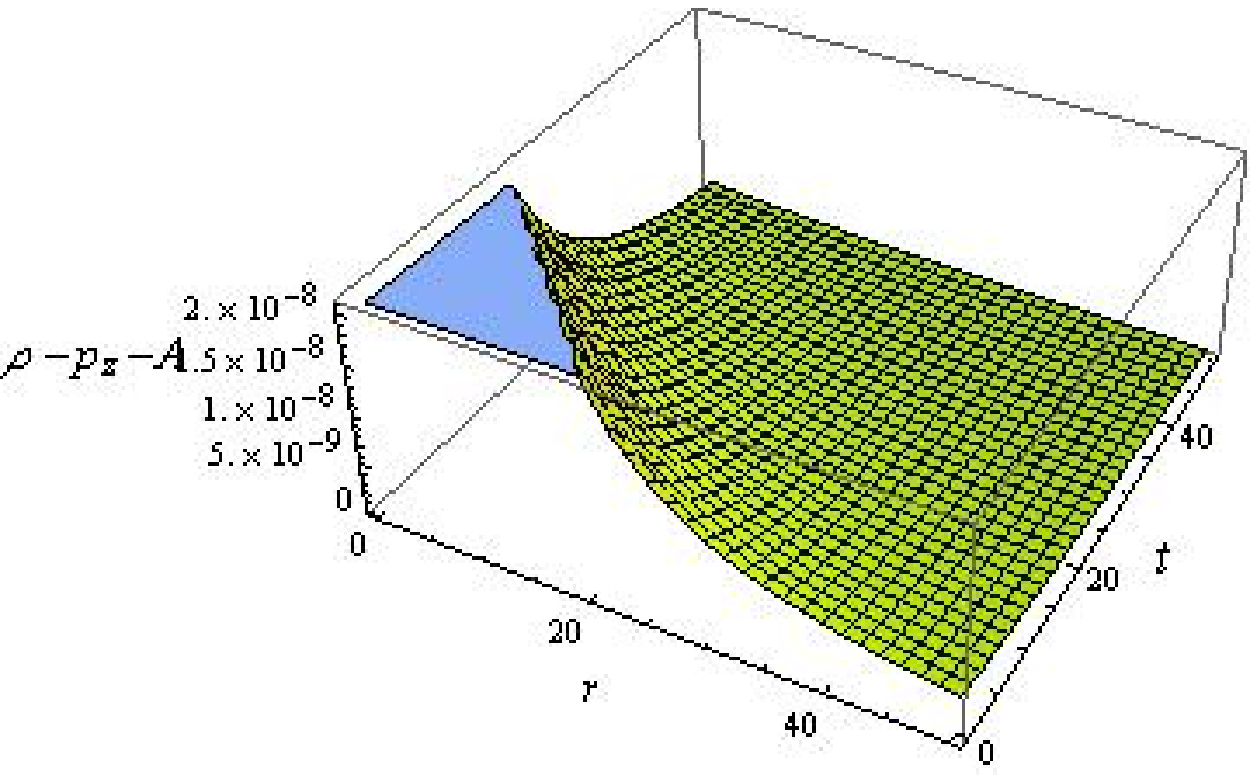,width=0.55\linewidth} \caption{Plots
for DEC for $\gamma=0.0001$, $\alpha=1$, $q=0.01$ and
$\lambda=-0.001$.}
\end{figure}

\section{Concluding Remarks}

Accelerated expansion of the universe is an observed phenomenon
which can affect astrophysical processes. To study the consequences
of the expanding universe on collapsing and expanding scenarios of a
stellar object, we consider charged anisotropic cylindrical source
in $f(R,T)$ framework. The solutions of Einstein-Maxwell field
equations governing the phenomena of collapse and expansion during
stellar evolution are discussed. We explore the role of
electromagnetic field and model parameter on the physical features.

In case of collapse solution, the expansion scalar, density,
pressures ($p_{r},~p_{\theta}$ and $p_{z}$), anisotropy and mass do
not change with radial coordinate. For expanding solution, the
change remains the same for both coordinates. The behavior of these
parameters with respect to time remains the same for both cases,
except $p_z$. The anisotropy is positive for both cases which
enhances the compactness of the system as discussed in \cite{Tf}. In
both cases, the increase in total charge has the same effects on
physical quantities. The effects of model parameter is different for
$p_{\theta}$ and $p_{z}$ in both cases while it is same for the
remaining quantities. We conclude that the collapse rate increases
for the collapse solution while the expansion rate decreases for the
expanding solution. It is found that the energy conditions are
satisfied in both cases showing physical viability of our solutions
for the considered values of constants.

Finally, we compare our results with those found in GR or
$\lambda=0$ \cite{8d}. For our collapse solution, the change in
physical quantities is related with increase in time not with radius
while in GR the quantities vary with radial coordinate but do not
vary with temporal one. In case of expanding cylinder, the physical
parameters vary with increase in both time and radius for our
solution while in GR the solution only induces a change with respect
to $t$. In both cases, the anisotropy decreases for our solutions
while it increases in GR. The increase in anisotropy can distort the
geometry of the system, i.e., solutions in GR can deform the shape
of the system. On the other hand, for our solutions the anisotropy
decreases leading to geometry preservation which is due to the dark
source terms. It is worthwhile to mention here that our solutions
satisfy the energy condition for chosen values of constants which
are not shown in the similar works \cite{8a}-\cite{8d}.

\vspace{0.5cm}

{\bf Acknowledgment}

\vspace{0.25cm}

We would like to thank the Higher Education Commission, Islamabad,
Pakistan for its financial support through the {\it Indigenous Ph.D.
5000 Fellowship Program Phase-II, Batch-III.}

\end{document}